\documentstyle[11pt,aaspp4]{article}

\def\etal{{\it et al.}}

\def\reff{r_{\rm eff}} 
\def\reffi{r_{\rm eff,i}} 
\def\reffj{r_{\rm eff,j}} 
\def\reffij{r_{\rm eff,i,j}} 
\def\reffK{r_{\rm eff,K}} 
\def\reffV{r_{\rm eff,V}} 
\def\reffopt{r_{\rm eff,opt}} 
\def\r14{$r^{1/4}$}

\def\meanmu{\langle\mu\rangle} 
\def\meanmuK{\langle\mu_K\rangle} 
\def\meanmuV{\langle\mu_V\rangle} 
\def\meanmuopt{\langle\mu_{\rm opt}\rangle} 
\def\meanmueff{\meanmu_{\rm eff}} 
\def\meanmueffi{\langle\mu_i\rangle_{\rm eff}} 
\def\meanmueffj{\langle\mu_j\rangle_{\rm eff}} 
\def\meanmueffK{\meanmuK_{\rm eff}} 
\def\meanmueffV{\meanmuV_{\rm eff}} 
\def\meanmueffopt{\meanmuopt_{\rm eff}} 
\def\mtot{m_{\rm tot}} 
\def\Vtot{V_{\rm tot}} 
\def\Ktot{K_{\rm tot}} 
\def\Dn{D_n} 
\def\DV{D_V} 
\def\DK{D_K} 
\def\Dopt{D_{\rm opt}} 

\def\Dnsigma{$\Dn$--$\sigma_0$}
\def\DVsigma{$\DV$--$\sigma_0$}
\def\DKsigma{$\DK$--$\sigma_0$}
\def\magarcsec2{mag~arcsec$^{-2}$} 
\def\rkc{$R_{\rm C}$}

\def\ikc{$I_{\rm C}$}

\def\mgtwo{Mg$_2$}

\def\mgtwosigma{\mgtwo--$\sigma_0$}
\def\meanSigma{\langle\Sigma\rangle_{\rm eff}} 
\def\meanSigmaK{\langle\Sigma_K\rangle_{\rm eff}} 

\slugcomment{To Appear in The Astronomical Journal, \\
	Vol. 116, 1998 October issue}

\lefthead{Pahre et al.}
\righthead{Physical Origins of the Fundamental Plane}

\begin{document}
 

\title{Near--Infrared Imaging of Early--Type Galaxies IV.  The Physical \\
	Origins of the Fundamental Plane Scaling Relations. }


\author{Michael A. Pahre\altaffilmark{1,2,3}, Reinaldo R. de Carvalho\altaffilmark{1,4}, and S. G. Djorgovski\altaffilmark{1}}
\altaffiltext{1}{Palomar Observatory, California Institute of Technology, 
	MS 105-24, Pasadena, CA \, 91125; \emph{email:}  george@astro.caltech.edu~. } 
\altaffiltext{2}{Present address:  Harvard-Smithsonian Center for Astrophysics,
60 Garden Street, MS 20, Cambridge, MA \, 02138; \emph{email:}  mpahre@cfa.harvard.edu~. } 
\altaffiltext{3}{Hubble Fellow.} 
\altaffiltext{4}{Departamento de Astrofisica, Observatorio Nacional, CNPq, Brazil; \emph{email:}  reinaldo@maxwell.on.br~.} 


\begin{abstract}
The physical origins of the Fundamental Plane (FP) scaling relations are investigated using
large samples of early--type galaxies observed at optical and near--infrared wavelengths.
The slope $a$ in the FP relation $\reff \propto \sigma_0^a \langle\Sigma\rangle_{\rm eff}^b$
is shown to increase systematically with wavelength from the $U$--band ($\lambda \sim 0.35 \mu$m)
through the $K$--band ($\lambda \sim 2.2 \mu$m).
A distance--independent construction of the observables is described which provides an
accurate measurement of the change in the FP slope between any pair of bandpasses.
The variation of the FP slope with wavelength is strong evidence of systematic variations in stellar 
content along the elliptical galaxy sequence, but is insufficient to discriminate between a 
number of simple models for possible physical origins of the FP.
The intercept of the diagnostic relationship between $\log \DK / \DV$ and $\log\sigma_0$ 
shows no significant dependence on environment within the uncertainties of the Galactic 
extinction corrections, demonstrating the universality of the stellar populations contributions
at the level of $\Delta (V-K) = 0.03$~mag to the zero--point of the global scaling relations.

Several other constraints on the properties of early--type galaxies---the slope
of the \mgtwosigma\ relation, the slope of the FP in the $K$--band, the effects of stellar populations 
gradients, and the effects of deviations of early--type galaxies from a dynamically homologous 
family---are included to construct an empirical, self--consistent model which provides a complete 
picture of the underlying physical properties which are varying along the early--type galaxy sequence.
The fundamental limitations to providing accurate constraints on the individual model parameters (variations
in age and metallicity, and the size of the homology breaking) appear to be subtle variations between among
stellar populations synthesis models and poorly constrained velocity dispersion aperture effects.
This empirical approach nonetheless demonstrates that there are significant systematic variations in both
age and metallicity along the elliptical galaxy sequence, and that a small, but systematic, breaking of 
dynamical homology (or a similar, wavelength independent effect) is required.
The intrinsic thickness of the FP can then be easily understood as small variations in age, metallicity,
and deviations from a homology at any particular point along the FP.
The model parameters will be better constrained by measurements of the change of the slope of the FP with
redshift; predictions for this evolution with redshift are described.
This model for the underlying physical properties that produce the FP scaling relations provides a 
comprehensive framework for future investigations of the global properties of early--type galaxies
and their evolution.
\end{abstract}




\section{Introduction}

It was immediately recognized by Dressler \etal\ (1987) and Djorgovski \& Davis (1987) that 
the existence of the bivariate Fundamental Plane (FP) correlations implied a strong regularity 
of the mass--to--light ratios ($M/L$) among elliptical galaxies.
They further noticed that the exact form of the dependence of the observable 
measuring the size of the galaxy (the half--light radius $\reff$) and the observable measuring 
the dynamics of the internal stellar motions (the velocity dispersion $\sigma$) required $M/L$ 
to vary slowly, but systematically, with the galaxy luminosity $L$.
If the virial theorem is combined with the assumptions that elliptical galaxies form a homologous
family and have a constant $M/L$, then the predicted dependence is $\reff \propto \sigma_0^2$. 
The observed power--law ``slope'' of this correlation (at optical wavelengths), however, ranges
from $\reff \propto \sigma_0^{1.2}$ to $\reff \propto \sigma_0^{1.4}$.
The difference between the predicted and observed correlations was taken as evidence that the
assumption of constant $M/L$ was in error:  elliptical galaxies would then have $M/L \propto L^{0.25}$.
The physical origin of this effect was unknown at the time, but later studies suggested it could 
be a result of variations in their stellar (\cite{renzini93}; \cite{ds93}; \cite{worthey96}; 
\cite{zepf96}; \cite{prugniel96}) or dark matter (\cite{ciotti96}) content.
The latter explanation, however, would be in contradiction to galactic wind models 
(\cite{arimoto87}) which can successfully account for the \mgtwo--$\sigma_0$ relation (\cite{ciotti96}).
Velocity anisotropy could also contribute to this effect (\cite{ds93}; \cite{ciotti96}) since
more luminous ellipticals tend to be more anisotropic (\cite{davies83}), but this effect
has not been explored in much detail.

The difference between the predicted and observed FP correlations might not be a result of
variations in the mass--to--light ratios among elliptical galaxies, but could instead be a 
systematic breakdown of homology along the galaxy sequence (\cite{capelato95}; \cite{pddc95}).
These deviations from a homologous family could take a structural form in that
the galaxies deviate from a pure de Vaucouleurs $r^{1/4}$ light profile:  if the distribution
of galaxy light follows a Sersic $r^{1/n}$ profile (\cite{sersic}), then a systematic
variation of $n$ as a function of luminosity could also be implied by the FP.
It is known that not all elliptical galaxies follow a strict $r^{1/4}$ light profile
(\cite{caon93}; \cite{burkert93}), but an investigation of galaxies in the Virgo cluster
suggests that this effect is not sufficient to explain fully the departure of the observed
FP correlations from the predictions (\cite{graham97}).
Alternatively, the breakdown of homology could be of a dynamical nature, in the sense that
the stellar velocity distributions vary systematically along the elliptical galaxy sequence.
This effect appears to follow directly from dissipationless merging (\cite{capelato95})
when the orbital kinetic energy of the pre--merger galaxies relative to each other is redistributed
into the internal velocity distribution of the merger product.
The FP correlations would be affected because the central velocity dispersion $\sigma_0$ then does
not map in a homologous way to the half--light velocity dispersion.
Since the global photometric parameters are typically evaluated at the half--light
radius, the exact details of the mapping of velocity dispersion from the core to the half--light
radius are essential.
An investigation of this effect on the FP by using velocity dispersion profiles from the 
literature suggests that it can contribute as much as one--half of the difference between the
observed optical FP correlations and the virial expectation assuming constant $M/L$
(\cite{busarello97}).

The purpose of the present paper is to explore in detail the properties of early--type
galaxies as a means of elucidating which underlying physical properties (and their systematic
variations within the family of early--type galaxies) are the origins of the FP and other global
correlations.
The two major classes of physical properties that will be explored here are stellar populations
(age and metallicity) and deviations from a homologous family.

The large catalogs of data used in the present paper will be summarized in \S\ref{fpmodels-data},
although more complete descriptions can be found in Pahre (1998b).
As will be shown in \S\ref{fpmodels-optre-kre}, the scaling radius changes {\sl systematically} 
from the optical to the near--infrared signifying the presence of color gradients, hence the 
use of $\reff(\lambda)$ in studying the variations of the FP with wavelength is an important
new element of this work.
The change of the slope of the FP between the optical and near--infrared bandpasses will be
described in \S\ref{fpmodels-opt-ir-diff} and \S\ref{fpmodels-opt-opt} using a methodology that is 
both distance independent and minimizes the cumulative effects of observational measurement 
uncertainties.
It will be shown in those sections that the slope of the FP increases systematically with wavelength.
The global, observed constraints on the properties of elliptical galaxies are enumerated in
\S\ref{fpmodels-general}, thereby providing a list of properties that any viable model for the 
origins of the FP must explain.
A detailed and self--consistent model will be constructed in \S\ref{fpmodels-selfconsistent} 
which simultaneously accounts for the changes of the slope of the FP with wavelength, the absolute 
value of the slope in the $K$--band, the \mgtwosigma\ relation, stellar populations gradients, and 
deviations of ellipticals from a dynamically homologous family.

\section{Description of the Data \label{fpmodels-data}}

The data used for this paper were compiled by Pahre (1998ab).
The photometric data are global parameters taken from recent surveys from the 
$U$ to $K$ bandpasses:  effective radii $\reff$, mean surface brightnesses $\meanmueff$ 
within those radii, total magnitudes $\mtot$, global colors, and the $\Dn$ parameter 
(\cite{dressler87}) defined in a self--consistent manner in all bandpasses.
All global photometric parameters (including in the $K$--band) were derived independently
from {\sl imaging} data with two exceptions:  the large survey of Faber \etal\ (1989) in the
$B$ and $V$ bands utilized photo-electric photometry to derive the parameters; and the study of
Prugniel \& Simien (1996) only provide $\reff$ in the $B$--band, but provide colors ranging
from $U$ to \ikc\ that were evaluated at the $B$--band $\reff$.
The catalogs used in the comparisons of photometry at near--infrared and optical bandpasses
can be found in Pahre (1998b), while the comparisons among optical bandpasses can be found in
Pahre (1998a).

The spectroscopic data are taken from the literature using the aperture correction methodology
of J\o rgensen \etal\ (1995b) and the small offsets between data sets derived by Smith \etal\ (1997).
The derived parameters are central velocity dispersion $\sigma_0$ and \mgtwo\ index, both
normalized to a physical scale of diameter $1.53 h_{75}^{-1}$~kpc corresponding to an angular diameter of
$3.4$~arcsec at the distance of the Coma cluster.

\section{Comparing Optical and Near--Infrared Effective Radii \label{fpmodels-optre-kre}}

\begin{figure}
	\epsscale{0.75}
	\plotone{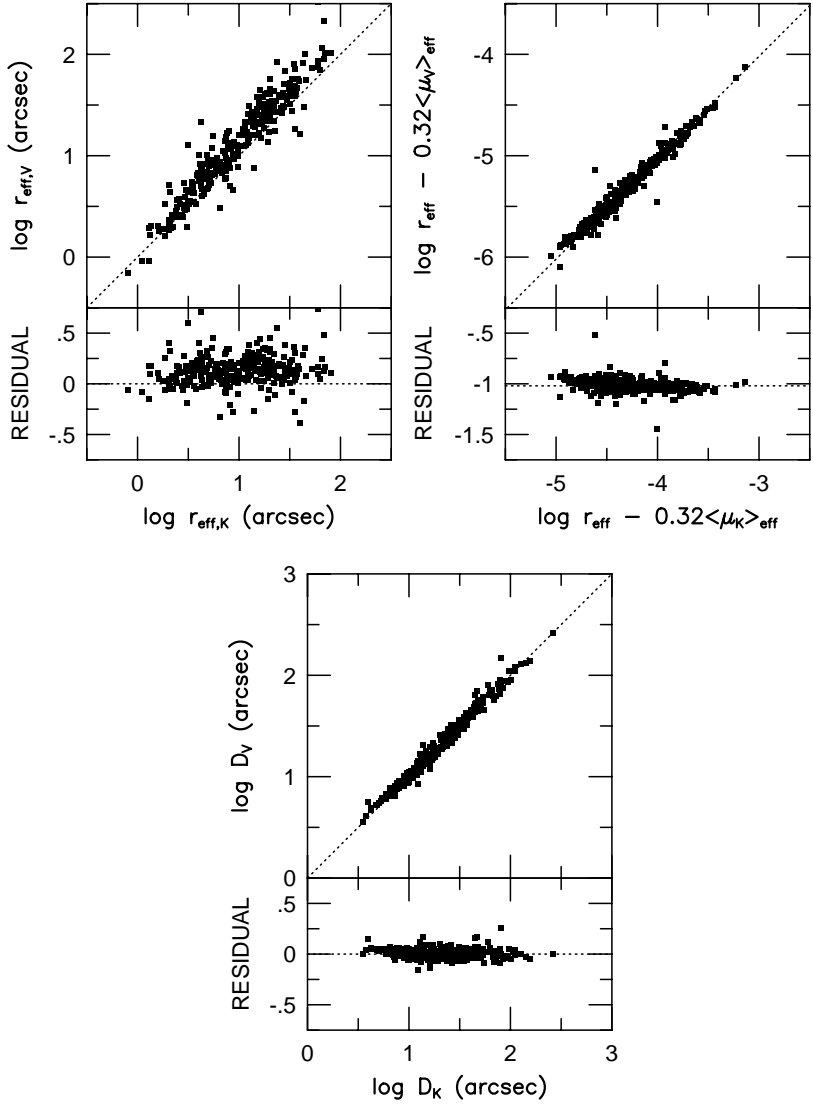}
	\caption[Comparison of the Effective Radii and Diam\-eters Derived From Optical and Near--Infrared Photometry]{
	Comparison between the estimates of the effective radius $\reff$ and diameter $D$ (i.e.,
	$D_V$ for the $V$--band, $D_K$ for the $K$--band).
	The quantity $\reff$ differs between the optical and near--infrared.
	The median offset for the comparison with the data of Lucey \etal\ (1991b, 1997) is
	$\log r_{\rm eff,V} - \log r_{\rm eff,K} = 0.08$~dex.
	This systematic variation is consistent with a color gradient of 
	$\Delta( \mu_V - \mu_K ) / \Delta \log \reff = -0.18$~mag~dex$^{-1}$, which is a good 
	match to the observed color gradients in $(B-R)$ and $(U-R)$ if their origin is in radial 
	variations in [Fe/H] (see text).
	}
	\label{fig-fpmodels-compare-reff}
\end{figure}

The comparison between radii and diameters, as measured independently in the optical and near--infrared,
are plotted in Figure~\ref{fig-fpmodels-compare-reff}.
There is a clear, systematic difference between $\log \reffK$ and $\log\reffopt$, in the
sense that the infrared effective radii are smaller than the optical ones.
The $\reffopt$ from the literature were not corrected for wavelength effects (as they should be due
to the presence of color gradients), hence it is necessary to use subsamples for individual filters to
make a meaningful comparison between optical and near--infrared effective radii.
For example, comparing $\reffV$ from Lucey \etal\ (1991b, 1997) with $\reffK$ shows a median offset of 
$\log \reffV - \log\reffK = +0.08$~dex ($N=94$).
This can be understood most simply as a change in scale length between the optical and near--infrared.
Using the formalism of Sparks \& J\o rgensen (1993), the change in scale length implied by a color gradient
is $\Delta s = \Delta \reff / \reff = 0.18$, which is converted into an isophotal color gradient of 
$\Delta ( \mu_V - \mu_K ) = \beta \Delta \log r$
with $\beta \sim \Delta s = -0.18$~mag~arcsec$^{-2}$~dex$^{-1}$.
This is similar to $\beta = -0.16$~mag~arcsec$^{-2}$~dex$^{-1}$ found by Peletier \etal\ (1990b), 
albeit for a small sample (12 galaxies) with a small FOV detector.
If the color gradient were due to a metallicity gradient, a simple change from [Fe/H]$ = 0$ to $-0.25$~dex
would produce a change in color of $\Delta(V-K) = 0.38$~mag, $\Delta(B-V) = 0.07$~mag, 
$\Delta(U-R) = 0.25$~mag, $\Delta(B-R) = 0.11$~mag (\cite{worthey94}).
Thus, a metallicity gradient of $\Delta $~[Fe/H]~$ / \Delta\log r = -0.12$ would be consistent with the
observed $\log \reffopt - \log\reffK = +0.08$~dex and the observed optical color gradients from the 
literature (\cite{sv78a}; \cite{franx89}; \cite{peletier90a}; \cite{jfk95a}).
If the trend of $(\log\reffopt - \log\reffK)$ with $\log\reffK$ is real, then this would also be
consistent with the size of the color gradients correlating with galaxy size and hence luminosity.

\placefigure{fig-fpmodels-compare-reff}

It is also apparent that there is no systematic offset between $\log\DV$ and $\log\DK$,
which is due to the good match between the assumed mean galaxy colors of $(V-K)=3.2$~mag
used in the definition of $\DK$ and the true mean galaxy color.
The rms of this difference is $0.048$~dex, which is significantly larger than the
uncertainties in both measurements added in quadrature.
Part of this effect is due to the change in the slope of the FP between the optical
and near--infrared which causes a correlation of $\log\DK - \log\DV$ with $\log\sigma_0$.
This effect will be discussed in \S\ref{fpmodels-opt-ir-diff}.

There is a systematic offset of the quantity $(\log\reffK - 0.32\meanmueffK) - (\log\reffV - 0.32\meanmueffV)$
in Figure~\ref{fig-fpmodels-compare-reff} which is primarily due to the mean color in $(V-K)$,
but some of this effect is also due to the correlation with $\sigma_0$ due to the
change in the slope of the FP between the optical and near--infrared.

\section{The Difference in Slope Between the Optical and Near--Infrared FP \label{fpmodels-opt-ir-diff}}

\subsection{The Traditional Method to Measure the Change in Slope of the FP \label{fpmodels-opt-ir-old} }

The near--infrared FP has been shown in Pahre \etal\ (1998) to be 
represented by the scaling relation
$\reff \propto \sigma_0^{1.53 \pm 0.08} \langle\Sigma_K\rangle_{\rm eff}^{-0.79 \pm 0.03}$.
This relation shows a significant deviation from the optical forms of the FP:
$\reff \propto \sigma_0^{1.24 \pm 0.07} \langle\Sigma\rangle_{\rm eff}^{-0.82 \pm 0.02}$
(\cite{jfk96}) and
$\reff \propto \sigma_0^{1.38 \pm 0.04} \langle\Sigma\rangle_{\rm eff}^{-0.82 \pm 0.03}$
(\cite{hudson97}) in the $R$--band; or
$\reff \propto \sigma_0^{1.13} \langle\Sigma\rangle_{\rm eff}^{-0.79}$ in the $V$--band
(\cite{guzman93b}).
A simple conclusion can be drawn from these data:  the slope of the FP increases with wavelength.
While the trend appears clear in all these comparisons, the statistical significance of any
one comparison is not overwhelming.
For example, the change in slope from $r$--band to $K$--band is $+0.29 \pm 0.11$, which
is at the $< 3 \, \sigma$~confidence level (CL).

\subsection{The New, Distance--Independent Method to Measure the Change in Slope of the FP \label{fpmodels-opt-ir-new} }

A more direct comparison will be made in this section between the optical and near--infrared
FP relations by explicitly fitting the difference in slope for only those galaxies in common between 
a given optical survey and the near--infrared survey {\sl using the same central velocity 
dispersions for both.}
This will provide reasonable estimates of both the change in slope and its 
uncertainty.\footnote{J\o rgensen \etal\ (1996) fit their data in $U$, $B$, and $g$ bandpasses by
assuming the cluster distances from the $r$--band solution.  While this is certainly an improvement
over the free fitting method because it offers an additional constraint on the problem, we consider
that the method which follows to be more elegant due to its independence from assumptions of distance.}

The method used here takes advantage of the observation that the quantity $(\log\reff - b \meanmueff)$
in the FP has a value of $b \sim 0.32$ that is independent of the wavelength studied or fitting method 
adopted.
Note, for example, that $b = 0.326 \pm 0.011$ for the $R_{\rm C}$--band (\cite{hudson97}),
$b = 0.328 \pm 0.008$ for the $r$--band (\cite{jfk96}), 
$b = 0.320 \pm 0.012$ for the $I_{\rm C}$--band (\cite{scodeggio97}), and
$b = 0.314 \pm 0.011$ for the $K$--band (\cite{pahre98kfp}).
This agreement occurs despite the different fitting methods employed in each 
study.\footnote{Prugniel \& Simien (1996) constructed a form of the FP (their Equation 3)
which is equivalent to the modified Faber--Jackson relation (\cite{pahre98kfp}) in our notation.
They argued that $\gamma$ in $(1+2\gamma)/(1+2\beta) = -b'$ (using their notation for $\beta$
and $\gamma$), where $2/(1+2\beta)=a'$, was poorly constrained but consistent with zero, and hence 
set $\gamma=0$.  This assumption causes the power law index $b'$ for the surface brightness to vary with 
$\beta$ and hence to vary with wavelength, {\sl in contradiction to all of the studies quoted in the text} 
(for a range in wavelengths).  }
By assuming $b \equiv 0.32$, the optical and near--infrared forms of the FP reduce to
\begin{equation}
	\begin{array}{rcl}
	\log \reffK ({\rm arcsec}) 		& = 	&\, \, a_K \log \sigma_0 + 0.32 \meanmueffK + c_{i,K} \\
	\log \reffopt ({\rm arcsec}) 		& = 	&\, \, a_{\rm opt} \log \sigma_0 + 0.32 \meanmueffopt + c_{i,{\rm opt}}
	\end{array}
	\label{fpmodels-eq-optfp-and-kfp}
\end{equation}
Taking the difference of these equations produces
\begin{equation}
	( \log\reffK - 0.32 \meanmueffK ) - ( \log\reffopt - 0.32 \meanmueffopt ) 
		= \Delta a \times \log\sigma_0 + c_{K,{\rm opt}}
	\label{fpmodels-eq-kfp-minus-optfp}
\end{equation}
where $\Delta a = a_{K} - a_{\rm opt}$ is the difference in FP slope between the near--infrared and optical,
and the constants have been combined into $c_{K,{\rm opt}}$.
There is no distance dependence on either the left--hand or right--hand sides of
Equation~\ref{fpmodels-eq-kfp-minus-optfp}, assuming that seeing corrections have been applied to
$\reff$ and $\meanmueff$, and aperture corrections to $\sigma_0$.
Therefore, all galaxies, both in clusters and the field, can be studied.
When the optical FP is subtracted from the near--infrared FP, as has been done for 
Equation~\ref{fpmodels-eq-kfp-minus-optfp}, then the left--hand side of the equation is related to
the mean color offset between the two bandpasses---albeit corrected for a small, but systematic, reduction in
effective radius from the optical to the near--infrared bandpasses.

Some literature sources measure a $\Dn$ parameter instead of $\reff$ and $\meanmueff$, hence an 
equivalent equation for the change in slope of the \Dnsigma\ relation is
\begin{equation}
	\log \DK - \log \Dopt	= \Delta a \times \log\sigma_0 + c_{K,{\rm opt}} .
	\label{fpmodels-eq-dk-minus-dv}
\end{equation}

All of the galaxies in the $K$--band sample with companion optical measurements of 
$\reff$ and $\meanmueff$ or $\Dn$ were fit to Equations~\ref{fpmodels-eq-kfp-minus-optfp} 
or \ref{fpmodels-eq-dk-minus-dv}, respectively.
The sum of the absolute value of the residuals was minimized in a direction orthogonal 
to the relation using the program GAUSSFIT (\cite{gaussfit}), and the uncertainties on 
$a$ and $c_{K,{\rm opt}}$ were determined from bootstrap resampling of the data.
This combined sample used mean colors to convert data in the $B$, $r$, $R_{\rm C}$, or
$I_{\rm C}$ bands into the $V$--band, with the exception of the Faber \etal\ (1989) data set,
for which the observed $(B-V)_0$ were used.
The problem with using mean colors is that no account is made for the color--magnitude 
relation between the observed and $V$ bandpasses, hence the slope of the FP does not
change {\sl by construction}.
For this reason, separate comparisons are made for each literature source for
those galaxies in common, but keeping the data in the original bandpass.
All of these fits are listed in Table~\ref{table-fpmodels-kfp-optfp}.
Note that the catalogs prepared for the purpose of this comparison have been put onto a common
extinction scale as described by Pahre (1998b).
Several outlier data points were excluded from these fits:  
D45 in Klemola~44, PER199 in Perseus, and D27 in Coma from the comparison with Faber \etal\ (1989);
E160G23 and NGC~4841A/B in Coma from the comparison with Scodeggio \etal\ (1997);
and NGC~6482 from all comparisons.

\placetable{table-fpmodels-kfp-optfp}

\begin{figure}
	\epsscale{0.55}
	\plotone{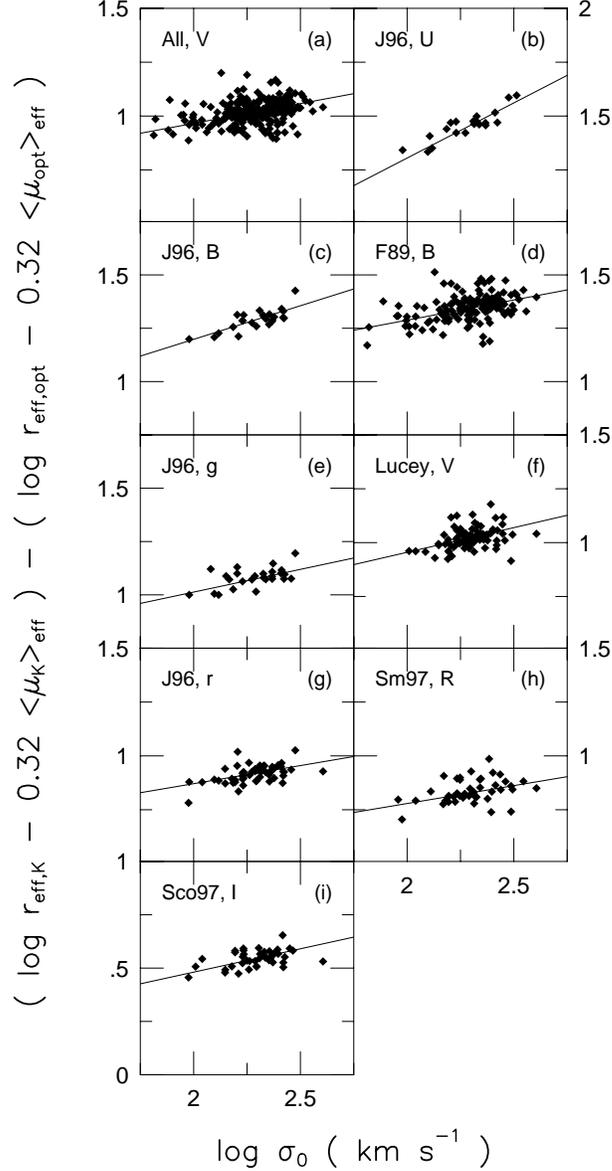}
	\caption[ Comparison of the Slope of the Optical and Near--Infrared Funda\-ment\-al Plane ]{
	Comparison of the slope of the optical and near--infrared FP relations.
	Plotted as the vertical axis is the difference in $\log\reff - 0.32\meanmueff$ 
	in the sense of $K$--band minus optical, while velocity dispersion is the horizontal axis.
	If the optical and near--infrared FP relations had identical slopes, then the points
	would lie on a horizontal line.
	The positive value of the linear regression in each case signifies a steepening of the
	FP as the wavelength moves from the optical to the near--infrared.
	The slopes and intercepts of these comparisons are listed in Table~\ref{table-fpmodels-kfp-optfp}.
	The literature comparisons are:  Faber \etal\ (1989; F89); 
	Lucey \& Carter (1988), Lucey \etal\ (1991a,b), and Lucey \etal\ (1997);
	J\o rgensen \etal\ (1996; J96); Smith \etal\ (1997; Sm97);  and Scodeggio \etal\ (1997; Sco97).
	}
	\label{fig-fpmodels-optfpkfp}
\end{figure}

\begin{figure}
	\epsscale{0.75}
	\plotone{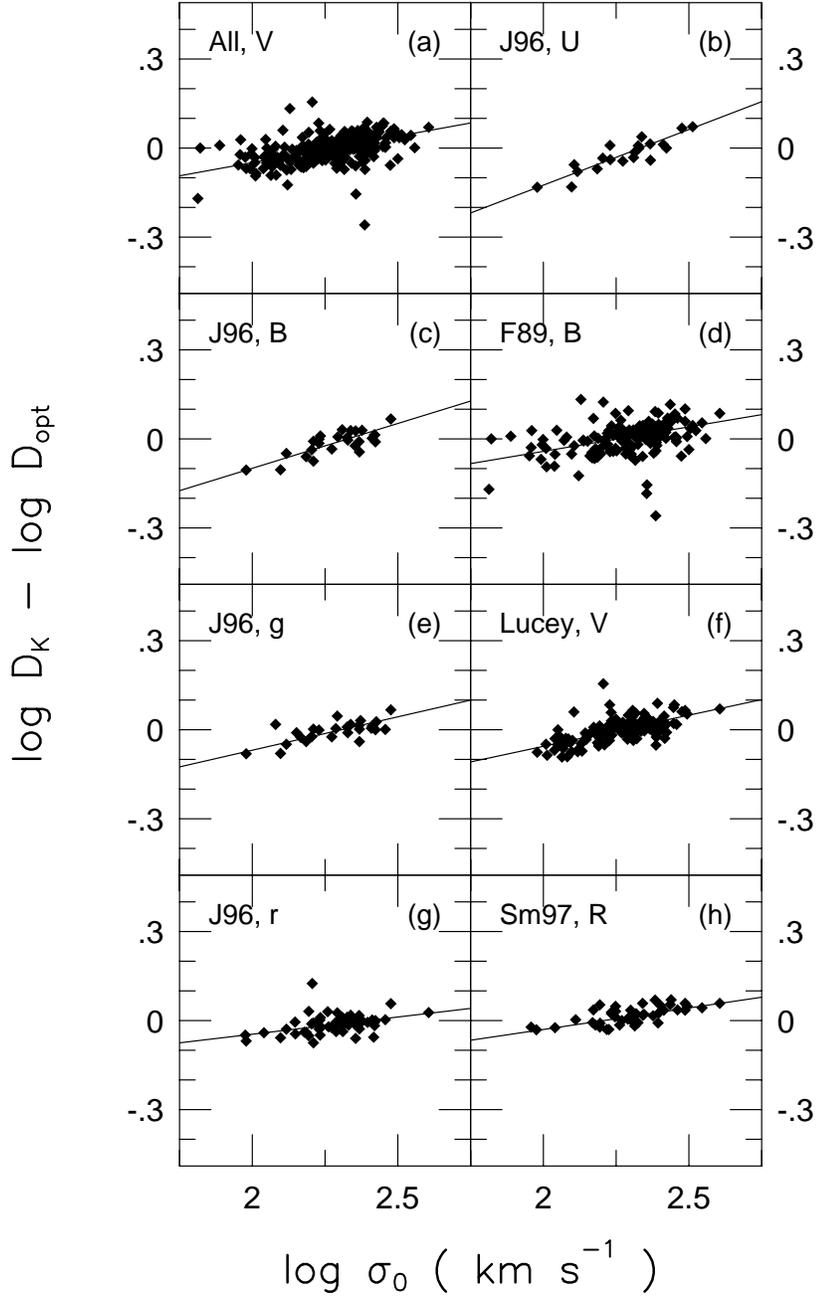}
	\caption[Comparison of the Slope of the Optical and Near--Infrared \Dnsigma\ Relation]{
	Comparison of the slope of the optical and near--infrared \Dnsigma\ relations.
	Plotted as the vertical axis is the difference in $\log\DK - \log\Dopt$, while
	velocity dispersion is the horizontal axis.
	If the optical and near--infrared FP relations had identical slopes, then the points
	would lie on a horizontal line.
	The positive value of the linear regression in each case signifies a steepening of the
	FP as the wavelength moves from the optical to the near--infrared.
	The literature comparisons are the same as in Figure~\ref{fig-fpmodels-optfpkfp}.
	}
	\label{fig-fpmodels-dvdk}
\end{figure}

The fits for each of the subsamples are plotted in 
Figures~\ref{fig-fpmodels-optfpkfp} and \ref{fig-fpmodels-dvdk} for the comparisons
of $\log\reff - 0.32\meanmueff$ and $\Dn$, respectively.
If the optical and near--infrared FP relations had the same slope, then all points would
lie on a horizontal line in these figures; this is clearly not the case.
The statistical significance of each regression for the comparison of $(\log\reff - 0.32\meanmueff)$
is at the 2--6~$\sigma$ confidence level (CL), while the significance for the $(\log\DK - \log\Dopt)$
comparison is at the 3--10~$\sigma$~CL.
As a demonstration of how the method adopted here is superior to the alternate method of fitting the optical
and near--infrared FP relations independently and then comparing their slopes, notice that the
$0.06$~dex uncertainty in Table~\ref{table-fpmodels-kfp-optfp} for the J\o rgensen \etal\ (1996) $r$--band 
subsample is nearly a factor of two smaller than the $0.11$~dex uncertainty derived 
when the independently fitted slopes were compared in \S\ref{fpmodels-opt-ir-old}, despite the fact that
one--fourth the number of galaxies were used in this newer method.
The difference is most likely due to several factors:  an identical sample of galaxies is
studied simultaneously in both the optical and near--infrared; the velocity dispersion term used is
identical for both FP relations; there is no assumption about the distance to a given galaxy or
its cluster; and the uncertainties in the velocity dispersion are not applied twice in the
estimation of uncertainties.

\placefigure{fig-fpmodels-optfpkfp}

\placefigure{fig-fpmodels-dvdk}

All of the optical to near--infrared comparisons of $(\log\reff - 0.32\meanmueff)$, with the
exception of the $U$--band comparison with J\o rgensen \etal\ (1995a), are
statistically indistinguishable.
The uncertainties, however, are large enough that small but real trends with
wavelength in the optical are not excluded.
While the \Dnsigma\ relations show no statistical difference between the $B$ and $V$ bandpasses,
there is a change in the slope between the optical to near--infrared data (significant at the
2--3~$\sigma$~CL) from the $V$ to the $R$ (or $r$) bandpasses, or from the $U$ to any other bandpass.
The reason that the $B$ band does not match this trend, or that the $(\log\reff - 0.32\meanmueff)$
comparison did not show the effect, is that these other comparisons have substantially larger
observational measurement uncertainties.
The $\Dn$ parameter can be measured 10--50\% more reliably than $(\log\reff - 0.32\meanmueff)$,
whether in the optical (\cite{jfk96}; \cite{smith97}; \cite{lucey97}) or near--infrared
(\cite{pahre98kfpdata}).

\subsection{Possible Environmental Effects on the FP  \label{kfp-opt-ir-environment}}

\begin{figure}
	\epsscale{0.5}
	\plotone{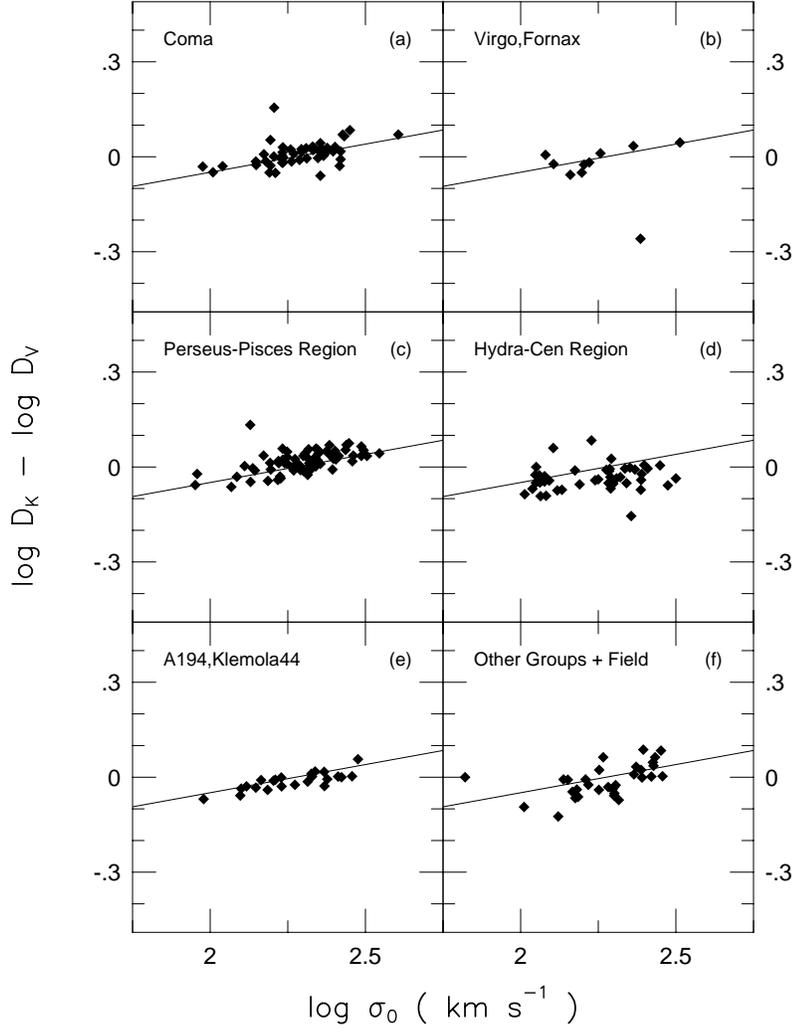}
	\caption[Comparison of the \DVsigma\ and \DKsigma\ Relations in Different Environments and Regions]{
	Comparison of the difference in slope and intercept of the \DVsigma\ and \DKsigma\ relations
	in various regions and density environments.
	The same straight line fit to the entire sample is displayed in each panel to ease comparison.
	Note how there is little or no difference between the slope or intercept in any panel and the
	mean relation.
	The largest possible offset is in the Hydra--Centaurus region, but this effect is smaller
	and {\sl of opposite sign} when compared to the effect found by Guzm\'an (1995).
	The large difference between Guzm\'an (1995) and the present work is fully explained by
	different assumptions of Galactic extinction in the Hydra--Centaurus region; the formal uncertainty on
	the IRAS 100$\mu$m to $A_B$ extinction conversion adopted here is consistent with there being no
	offset between the mean relation and the Hydra--Centaurus galaxy subsample.
	This figure provides evidence that there are no significant environmental effects on the \Dnsigma\
	relation.
	}
	\label{fig-kfp-dvdk-byregion}
\end{figure}

A significant offset in the relationship between $(\log\DK - \log\DV)$ and $\log\sigma_0$, as
measured separately in the Coma cluster and the Hydra--Centaurus Region, was identified by
by Guzm\'an (1995).
If this were the case, it would suggest that there are significant environmental effects on
the elliptical galaxy correlations, thereby preventing their utility as accurate distance
indicators.
The present paper includes larger samples of galaxies both in Coma and Hydra--Centaurus, as
well as other rich clusters and low density environments, hence this effect can be re--analyzed.
The data were broken down into six regions of the sky or similar density environments,
compared to the overall solution 
(as listed in Table~\ref{table-fpmodels-kfp-optfp}), and are displayed in Figure~\ref{fig-kfp-dvdk-byregion}.

\placefigure{fig-kfp-dvdk-byregion}

Guzm\'an (1995) found that $(\log\DK-\log\DV)$ at a given $\log\sigma_0$ was $\sim 0.05$~dex \emph{larger} 
in Hydra--Centaurus region than in Coma, but panel (d) of Figure~\ref{fig-kfp-dvdk-byregion} shows that
it is $< 0.03$~dex \emph{smaller}.
Most of the difference between the results of Guzm\'an and the present work can be explained
by different assumptions of Galactic extinction:  he apparently used Burstein \& Heiles (1982) 
maps (which are based on galaxy number counts and neutral gas emission) as the estimator of $A_B$, while
this work uses the 100$\mu$m emission (as measured by IRAS) as the estimator of $A_B$.
While the two estimates agree fairly well in a global sense, they disagree by 0.1--0.2~mag
in Hydra--Centaurus, in the sense that Burstein \& Heiles (1982) underestimates $A_B$.
Nonetheless, the formal error on $A_B$ due to the uncertainties of the IRAS conversion
in Laureijs, Helou, \& Clark (1994), are sufficient to bring all the galaxies in this
region into agreement with the global relation between $(\log\DK-\log\DV)$ and $\log\sigma_0$.
Furthermore, the somewhat larger scatter of the galaxy properties found in this region, compared 
to the cluster subsamples in the other panels, also argues for significant, patchy dust extinction.
It is interesting to note that by increasing $A_B$ in Hydra--Centaurus, the distance estimations
of Lynden--Bell \etal\ (1988) would place these galaxies closer to us, thereby strengthening the
statistical evidence for a Great Attractor causing bulk motions of the galaxies in this region.

Inspection of Figure~\ref{fig-kfp-dvdk-byregion} leads to the conclusion that
there is no evidence that either the slope or intercept of the elliptical galaxy correlations 
are dependent on environment.
For example, the difference in intercepts between the field and group sample ($N=32$) and
all other galaxies ($N=213$; both using a $\pm 0.15$~dex clip to exclude outliers) in 
Figure~\ref{fig-kfp-dvdk-byregion} is only $0.010 \pm 0.009$~dex, which corresponds to a
difference of $0.03 \pm 0.03$~mag in $(V-K)$ between the two sub-samples (in the sense that
the field galaxies are marginally bluer).
Using the Bruzual \& Charlot (1996) models, this offset in $(V-K)$ color constrains the
difference in mean age between field and cluster early--type galaxies to be $\leq 0.045$~dex.
The extinction correction $E(V-K) = 0.65 A_B$, so uncertainties in $A_B$ of $\geq 0.05$~mag
(which corresponds to the $\Delta (V-K) = 0.03$~mag effect on the intercept)
are obtained for $A_B \geq 0.3$~mag in the Laureijs \etal\ (1994) extinction estimates.
Extinction estimates for the Hydra--Centaurus region, Perseus, and Pegasus exceed this value,
while high latitude clusters such as Coma, Abell~2199, Klemola~44, and Fornax have little
or no extinction; the field and group sample shows a wide range in extinction
$0.01 \geq A_B \geq 0.84$~mag (\cite{pahre98kfpdata}).
Uncertainties in foreground Galactic extinction therefore appear to be the dominant source
of uncertainty contributing to any possible differences between field and cluster early--type
galaxies measured using this method.
The difference in FP intercept and scatter between galaxies in clusters and the general field 
found by de Carvalho \& Djorgovski (1992) would be naturally explained by errors in estimating 
dust extinction (which also affects the present paper) or estimating distances (which does not 
affect the present paper), and not by any intrinsic differences in the stellar populations 
among elliptical galaxies that correlates with their environment.
Difficulties in estimating Galactic dust extinction therefore appear to be the limiting factor 
for \emph{optical} distance scale work using the elliptical galaxy correlations.
The sample used here is also small (only $\sim 30$ galaxies in loose groups and the general
field), hence these results naturally require confirmation with larger galaxy samples.

While stellar populations variations with environment appear to be excluded by the results in
Figure~\ref{fig-kfp-dvdk-byregion}, the effects due to variations in homology breaking or dark 
matter content with environment would cancel out in this figure and analysis.
Hence, environmental variations that appear as a wavelength independent property, such as
homology breaking or systematic variations in dark matter content, are not constrained by
this analysis.
A test for these effects could entail comparing the slope of the FP, as measured separately
at $V$ and $K$ (as in \S\ref{fpmodels-opt-ir-old}), between clusters and the field.
Unfortunately, such a test will suffer from systematic errors due to distance which were 
eliminated in the approach advocated in \S\ref{fpmodels-opt-ir-new}.
The use of an accurate distant indicator (such as surface brightness fluctuations; \cite{tonry97})
could circumvent this problem; such an analysis will be reserved for a future contribution.

In summary, while there appear to be no variations in the intercept of the
FP with environment due to stellar populations effects, this result is uncertain at the
level of the correction for Galactic extinction and subject to the assumption
that the slope of the FP is not varying with environment.
The issue of whether there exists variations in homology breaking or dark matter content
with environment is left open.

\section{Comparing the Fundamental Plane Among Various Optical Bandpasses  \label{fpmodels-opt-opt}}

When comparing the optical and near--infrared FP in \S\ref{fpmodels-opt-ir-diff}, there were hints 
that the differences might be larger for the $U$--band than for, say, the \rkc--band.
Small changes of the slope of the FP between $U$ and $r$, for example, were reported by 
J\o rgensen \etal\ (1996) and Djorgovski \& Santiago (1993) in the sense that the $U$--band FP 
has the shallowest slope.
Although the differences determined in this manner were of small significance, it should
be possible to improve the significance by removing the distance assumptions and using the
method described above.
Catalogs of optical global photometric parameters and velocity dispersions were compiled from the 
literature in the same manner as for the near--infrared catalogs (see \cite{pahre98kfpdata}).
An additional catalog consisting of $(U-B)$, $(B-V)$, $(V-R_{\rm C})$, and $(V-I_{\rm C})$ colors and
velocity dispersion was constructed from Prugniel \& Simien (1996) without modification.
Since Prugniel \& Simien do not measure $\reff$ and $\meanmueff$ independently for each of the five
bandpasses, the differences in $\log\reff - 0.32 \meanmueff$ were taken to be $0.32$ multiplied by
the color.
This approach is only partially correct since it does not account for the presence of color gradients,
but the systematic errors resulting from this simplification are small.

\placetable{table-fpmodels-optfp-optfp}

\begin{figure}
	\epsscale{0.6}
	\plotone{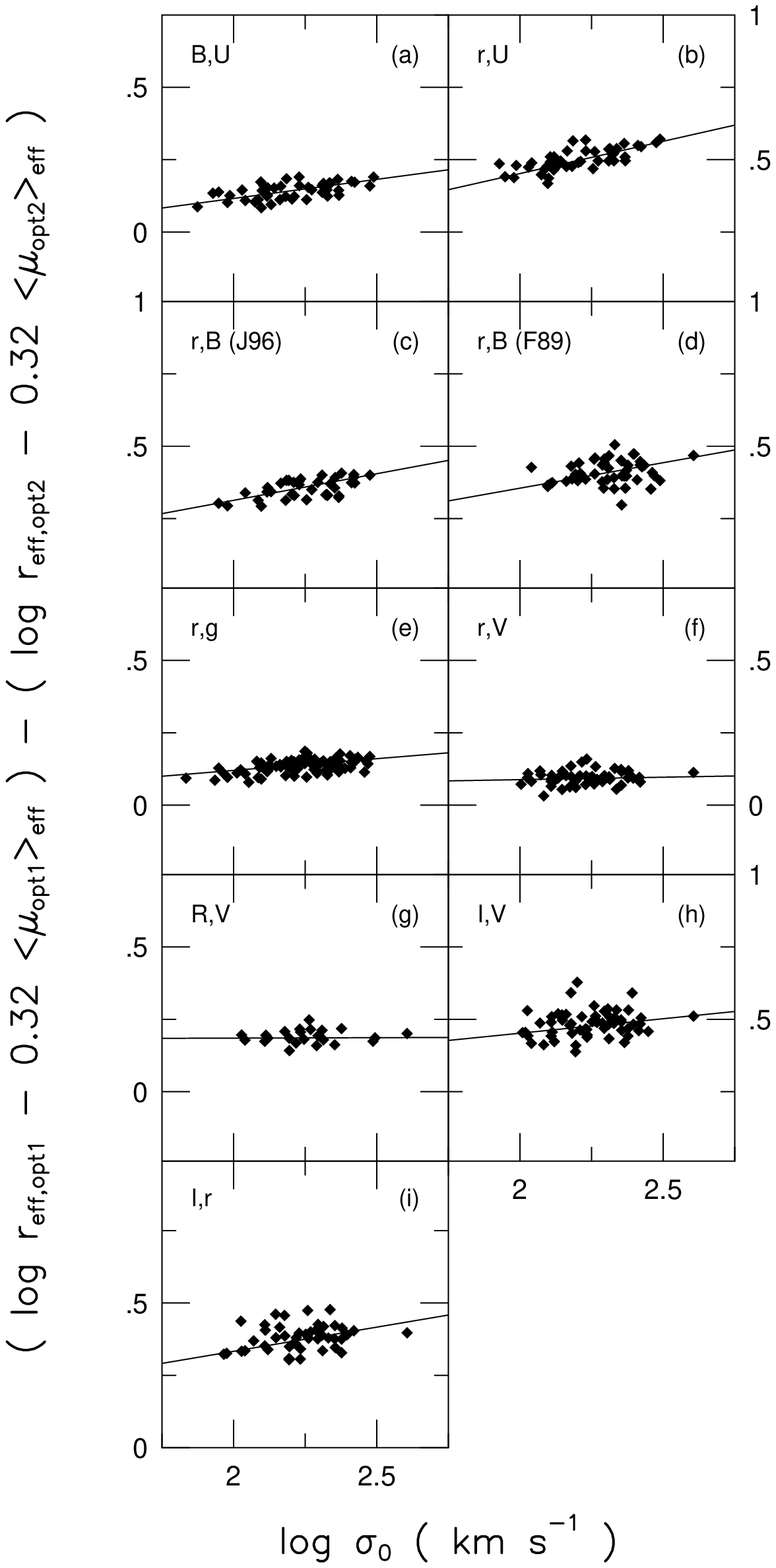}
	\caption[Comparison of the Slope of the Fundamental Plane Among Optical Band\-pass\-es]{
	Comparison of the slope of the FP relations among various optical bandpasses from
	$U$ to \ikc.
	In each panel, the vertical axis is the difference in $\reff - 0.32\meanmueff$ measured
	in each of the pair of bandpasses; the bandpasses are identified in the upper--left
	corner of each panel, and the difference is in the sense of the first bandpass minus
	the second.
	The FP slope is steeper in redder bandpasses as is evidenced by the positive correlation
	in nearly every panel.
	Literature sources and regressions are taken from Table~\ref{table-fpmodels-optfp-optfp}.
	}
	\label{fig-fpmodels-optfpoptfp}
\end{figure}

\begin{figure}
	\epsscale{0.75}
	\plotone{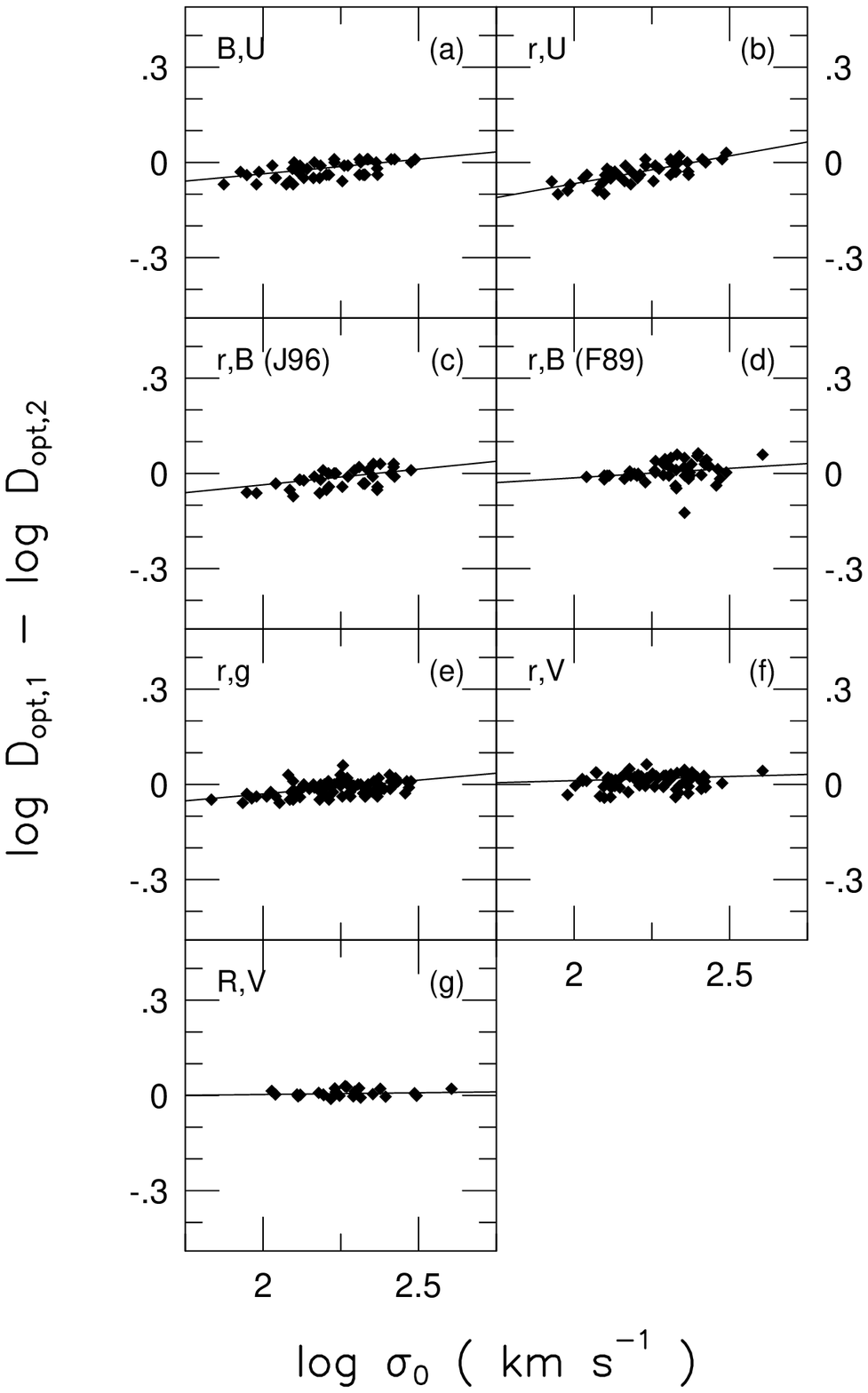}
	\caption[Comparison of the Slope of the \Dnsigma\ Relation Among Optical Band\-pass\-es]{
	Comparison of the slope of the \Dnsigma\ relation among various optical bandpasses from
	$U$ to \ikc.
	In each panel, the vertical axis is the difference in $\log\Dn$ measured
	in each of the pair of bandpasses; the bandpasses are identified in the upper--left
	corner of each panel, and the difference is in the sense of the first bandpass minus
	the second.
	The slope of the \Dnsigma\ relation is steeper in redder bandpasses as is evidenced by 
	the positive correlation in nearly every panel.
	Literature sources and regressions are taken from Table~\ref{table-fpmodels-optfp-optfp}.
	}
	\label{fig-fpmodels-doptdopt}
\end{figure}

\begin{figure}
	\epsscale{1.0}
	\plotone{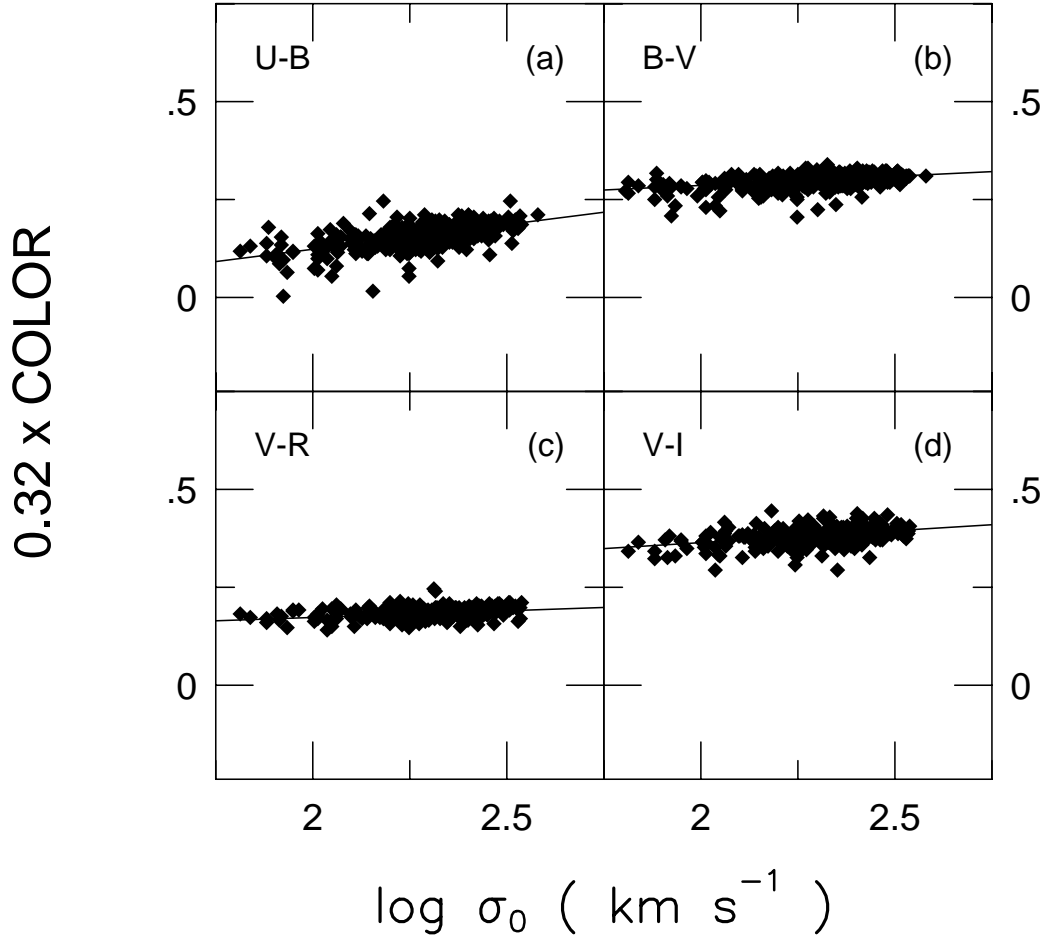}
	\caption[Comparison of the Slope of the FP Among Various Optical Bandpasses Using the Color
		Information From Prugniel \& Simien (1996)]{
	Approximate comparison of the slope of the FP relations between various optical bandpasses
	using the color information from Prugniel \& Simien (1996).
	Since those authors did not measure $\reff$ and $\meanmueff$ independently for each bandpass,
	the quantity 0.32 times the color was substituted for the difference in $\reff - 0.32\meanmueff$
	between each pair of bandpasses.
	This approach is reasonably similar to Figure~\ref{fig-fpmodels-optfpoptfp}, although it does not fully account for
	the effects of color gradients on the slope of the FP.
	Regressions are taken from Table~\ref{table-fpmodels-optfp-optfp}.
	Notice that all panels show a regression with positive slope, once again indicating that
	the slope of the FP steepens with wavelength.
	\label{fig-fpmodels-optfpoptfp-colors}
	}
\end{figure}

All comparisons demonstrate that the redder bandpass has a steeper slope for the FP
as evidenced by a positive $\Delta a$, although in several cases $\Delta a$ is statistically
indistinguishable from zero.
The comparisons derived from surface photometry are displayed in Figures~\ref{fig-fpmodels-optfpoptfp} 
and \ref{fig-fpmodels-doptdopt}, while those derived from color information alone (i.e., \cite{prugniel96})
are displayed in Figure~\ref{fig-fpmodels-optfpoptfp-colors}.

\placefigure{fig-fpmodels-optfpoptfp}

\placefigure{fig-fpmodels-doptdopt}

\placefigure{fig-fpmodels-optfpoptfp-colors}

In several comparisons where the two bandpasses differ only slightly in wavelength---such as between
the $r$--band and $V$--band---there is no significant variation of the
slope of the FP.  In most cases, however, there is a statistically
significant 3--13 $\sigma$~CL positive regression, and in no case is there a negative
regression, in the analysis listed in Table~\ref{table-fpmodels-optfp-optfp}.

\section{General Constraints from the Elliptical Galaxy Scaling Relations  \label{fpmodels-general} }

The preceding section, and a number of earlier papers, describe a series of global properties of 
early--type galaxies that are elucidated from the exact forms of the scaling relations in various bandpasses.
These can be summarized as:
\begin{enumerate}
\item Early--type galaxies are well--described by a Fundamental Plane corresponding to 
	the scaling relation $\reff \propto \sigma_0^{1.53\pm 0.08} \meanSigmaK^{-0.79 \pm 0.03}$ 
	(\cite{pahre98kfp}).
\item As has been shown in \S\ref{fpmodels-opt-ir-diff}, the slope of the FP (the exponent for the $\sigma_0$ term)
	steepens significantly between the optical and near--infrared.   As shown in \S\ref{fpmodels-opt-opt}, 
	the slope of the FP steepens with wavelength even among the optical bandpasses.
\item The slope of the FP at all wavelengths is inconsistent with the relation
	$\reff \propto \sigma_0^{2} \meanSigma^{-1}$ which is expected from the virial theorem 
	under the assumptions of constant mass--to--light ratio and homology within the family 
	of elliptical galaxies.
\item The FP and \mgtwosigma\ relations may be thin, but they have significant, resolved intrinsic 
	scatter which cannot be explained by the observational uncertainties and does not have 
	a clear correlation with any particular indicator of metallicity or age (\cite{jfk96};
	\cite{pahre98kfp}).
\item The effective radius of early--type galaxies was shown in \S\ref{fpmodels-optre-kre} to be 
	systematically smaller at longer wavelengths, which is basically equivalent to the existence
	(and size) of color gradients in these galaxies if they result from metallicity gradients
	(\cite{peletier93}).
\end{enumerate}
There are several other relevant properties of early--type galaxies that can be added to the above list 
but were not directly shown in this paper:
\begin{enumerate}
\setcounter{enumi}{5}
\item The velocity dispersion measured in an aperture decreases with the increasing size of this aperture
	according to a power law; the exponent of this power law appears to show a correlation with
	the luminosity or size of the galaxy (\cite{jfk95b}; \cite{busarello97}).
\item The ratio of magnesium to iron appears to be over-produced in early--type galaxies relative to 
	the solar value (\cite{worthey92}), albeit with a significant spread in [Mg/Fe], implying the
	importance of type~II supernovae chemical enrichment and rapid massive star formation in the
	galaxy formation process, particularly for the most luminous elliptical galaxies.
\item The correlation of \mgtwo\ with $\sigma_0$ implies a connection between the chemical
	enrichment of a galaxy and the depth of its potential well.
\item Optical and near--infrared color gradients in elliptical galaxies imply isophotal populations gradients
	of the order $0.16$ to $0.30$~dex in [Fe/H] (or 1.5 times this in $\log$~age) per decade of 
	radius (Franx \etal\ 1989; Peletier \etal\ 1990a,b; \cite{peletier93}).
\item There is no known correlation between the size of the measured color gradient and the luminosity of
	the host galaxy (Peletier \etal\ 1990a), although some of the smallest galaxies show no gradients
	altogether.
\end{enumerate}
Any viable model to explain the global properties of early--type galaxies must be able to account for
all of these effects.

\section{A Self--Consistent Model for the Underlying Physical Parameters Which Produce the FP 
	Correlations  \label{fpmodels-selfconsistent} }

In this section, a series of models will be constructed and explored in order to determine 
if all the observational constraints in \S\ref{fpmodels-general} can be explained in a fully 
consistent manner.

\subsection{Modeling the Changes in the Slope of the FP Between Bandpasses \label{fpmodels-self-slope} }

The effects on broadband color of the change in slope of the FP with wavelength can be expressed 
in a simple manner.
Starting with the definition of total magnitude for a de Vaucouleurs profile,
\begin{equation}
	\mtot = -5 \log \reff + \meanmueff - 2.5 \log 2 \pi
\label{fpmodels-eq-mtot-def}
\end{equation}
and the definition of the change in slope $\Delta a_{j,i}$ of the FP from bandpass $j$ to bandpass $i$
(Equation~\ref{fpmodels-eq-kfp-minus-optfp}), the change in the global color $\Delta C_{i,j}$ between 
bandpass $i$ and $j$ is then
\begin{equation}
\begin{array}{rcl}
\Delta C_{i,j} 	& = & 3.125 \left[ \left( \log\reffj - 0.32 \meanmueffj \right) 
			- \left( \log\reffi - 0.32 \meanmueffi \right) \right] \\
		&   &   - 1.875 \left( \log\reffi - \log\reffj \right) \\
		& = & 3.125 \Delta a_{j,i} \Delta\log\sigma_0 - 1.875 \Delta\reffij
\end{array}
\label{fpmodels-eq-color-change}
\end{equation}
where $\Delta\reffij$ is the change of $\log\reff$ from bandpass $i$ to bandpass $j$,
and $\Delta\log\sigma_0 \sim 0.6$~dex is the change in $\log\sigma_0$ from one end of the FP 
to the other (a range within which $>90$\% of the galaxies lie).
The two terms on the right--hand side in Equation~\ref{fpmodels-eq-color-change} show the
effects of the change in slope of the FP and the presence of color gradients, respectively.

In multi--color studies of isophotal color gradients in elliptical galaxies, Peletier \etal\ (1990a,b)
and Franx \etal\ (1989) found consistent results if the underlying cause were metallicity gradients
of $-0.20$, $-0.16$, and $-0.3$~dex, respectively.
A simple stellar populations model can then be used to convert these estimates
into any broadband isophotal color gradient $\beta$ between $U$ and $K$.
The conversion from isophotal color gradient to $\Delta \log \reff$ is accomplished using Equation~21 of
Sparks \& J\o rgensen, such that $\Delta \log\reff = \beta / ( 2.3 \times 1.20)$.
Hence, only one parameter to represent the global mean metallicity gradient is introduced into
the sets of equations described by Equation~\ref{fpmodels-eq-color-change} for the 22 observed $\Delta a_{j,i}$
from Tables~\ref{table-fpmodels-kfp-optfp} and \ref{table-fpmodels-optfp-optfp}.

\begin{figure}
	\epsscale{0.75}
	\plotone{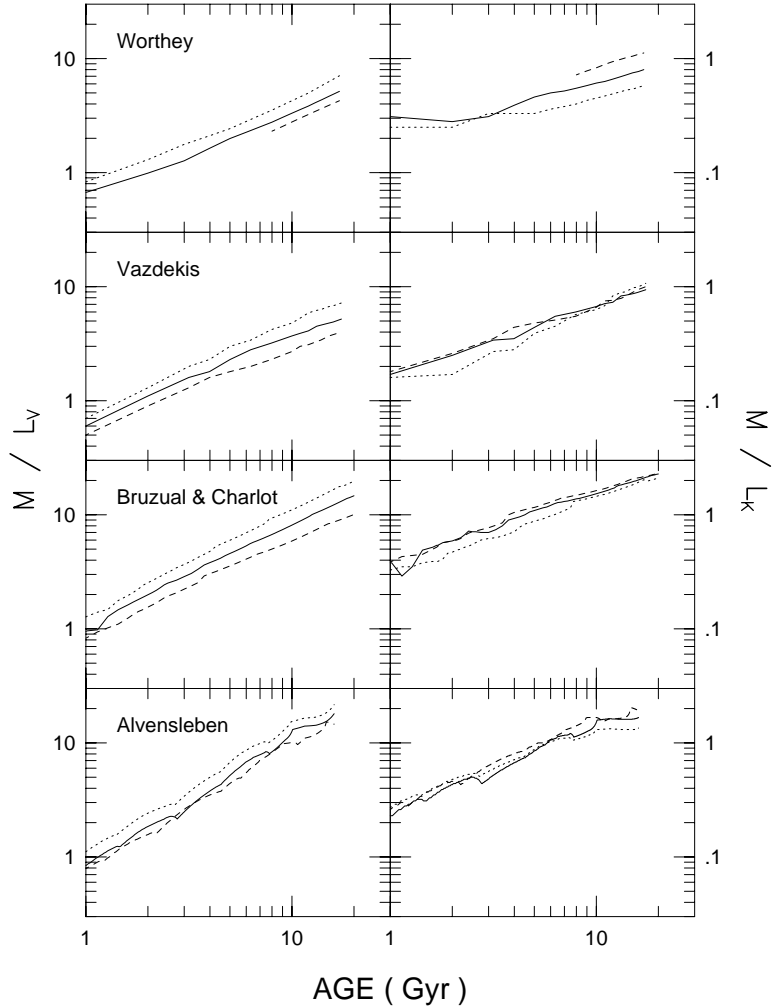}
	\caption[Comparison of $M/L_V$ and $M/L_K$ for Four Different Stellar Populations Models]{
	Comparison of $M/L_V$ (left) and $M/L_K$ (right) for four different stellar populations models
	from Worthey (1994), Vazdekis \etal\ (1996), Bruzual \& Charlot (1996, in preparation; as
	provided in \cite{leitherer96}), and Fritze-V. Alvensleben \& Burkert (1995).
	Iron abundances of [Fe/H]$= -0.4$ (dashed line), $0.0$ (solid line), and $+0.4$~dex (dotted line)
	are shown, except for the \cite{alvensleben95} models which have the $+0.3$~dex model 
	substituted for $+0.4$~dex.
	All models show similar variations in $M/L_V$ with both time and abundance, but not $M/L_K$.
	In particular, the Worthey (1994) models show a dependence of $M/L_K$ on [Fe/H]---such 
	that metal--rich systems have small $M/L_K$---while the Vazdekis \etal\ and Bruzual \& Charlot
	models have $M/L_K$ independent of metal abundance.
	}
	\label{fig-fpmodels-moverl-v-k}
\end{figure}

There is a significant difference in $M/L$ even among the most sophisticated of simple stellar
populations models (\cite{chwobr96}).
Nonetheless, use of such models in a {\sl differential} sense shows far less variation among
the models.
An example of this is given in Figure~\ref{fig-fpmodels-moverl-v-k}, where the mass--to--light ratio 
in the $V$--band and $K$--band is compared for four such models.
For large ages $t \geq 10$~Gyr, both the Vazdekis \etal\ (1996) and the
Bruzual \& Charlot (1996, as provided in \cite{leitherer96}) models show similar behavior
with $M/L_K$ independent of [Fe/H], while the Worthey (1994) models have $M/L_K$ {\sl inversely}
dependent on [Fe/H] and the Fritze-V. Alvensleben \& Burkert (1995) models are inconclusive.
In fact, Charlot, Worthey, \& Bressan (1996) showed in detail how three models differ strongly
in their near--infrared properties.
From the inspection of the near--infrared portion of Figure~\ref{fig-fpmodels-moverl-v-k}, 
we have chosen only to make detailed comparisons with the Vazdekis \etal\ and 
Bruzual \& Charlot models.

\placefigure{fig-fpmodels-moverl-v-k}

An additional question could be posed based on Figure~\ref{fig-fpmodels-moverl-v-k}:  given that there
are significant spreads in $M/L$ between the four models at any wavelength, are the {\sl changes} 
in $M/L$ (by varying the age and/or metal abundance) more consistent between the models?
This was addressed by Charlot \etal\ (1996) who showed that the variations among
three models was of order
$\delta (B-V) / \delta t \sim 0.004$~mag~Gyr$^{-1}$,
$\delta (V-K) / \delta t \sim 0.015$~mag~Gyr$^{-1}$, and
$\delta (M/L_V) / \delta t \sim 0.1$~M$_\odot$~L$_\odot^{-1}$~Gyr$^{-1}$ at $t \sim 10$~Gyr.
Hence a given model can be used to measure differential age or metallicity effects for an old 
stellar population while not providing an accurate absolute measure of either quantity.

The variations in magnitude as a function of changing [Fe/H] from $-0.4$~dex to $+0.4$~dex at $t = 11$~Gyr,
and separately as a function of changing age from 2 to 17~Gyr (at intervals of 1~Gyr) at [Fe/H]~$=0$~dex,
were calculated using the Bruzual \& Charlot models for the $UBVR_{\rm C}I_{\rm C}K$ bandpasses.
The same calculations were made for the Vazdekis \etal\ (1996) models.
For the modeling below, the Gunn $r$--band will be assumed identical (for differential effects) to the
Cousins \rkc--band, the Gunn $g$--band will be assumed identical to the $V$--band, and the $K_s$--band
assumed identical to the $K$--band.
These calculations are summarized in Table~\ref{table-fpmodels-bc96-ubvrik}.

\placetable{table-fpmodels-bc96-ubvrik}

\subsection{Additional Equations of Constraint \label{fpmodels-self-additionaleq} }

The fit to the \mgtwosigma\ relation (\cite{pahre98kfp}) provides an additional 
equation of constraint derived from the Bruzual \& Charlot models (using the variations specified
in Table~\ref{table-fpmodels-bc96-ubvrik}), namely
\begin{equation}
0.173 \pm 0.010 = \left[ 0.174 \Delta\log t 
	+ 0.278 \left( \Delta {\rm~[Fe/H]} 
		+ \left( {\Delta {\rm~[Fe/H]} \over 1.2 \times 1.6 \Delta\log r } \right) \right) \right] 
	\Delta\log\sigma_0
\label{fpmodels-eq-mg2sigma-constraint}
\end{equation}
The factor of 1.2 in the denominator converts the isophotal gradients into linear changes in
$\reff$ (from Equation~21 of \cite{sparks93}), while the factor of 1.6 converts this change in
$\reff$ into an aperture populations gradient (\cite{sparks93}, Equation~18).
This latter point is essential to recognize since \mgtwo\ is typically measured in an aperture 
of fixed physical size---and has been corrected to a fixed physical size using the methodology 
of J\o rgensen \etal\ (1995b).

In all cases, an isophotal populations gradient of $\beta = \Delta(\mu_i - \mu_j) / \Delta \log r$
between bandpasses $i$ and $j$ is converted to an equivalent change in effective radius
$\Delta \reffij = \beta / ( 1.2 \times 2.3 )$, where the factor of 2.3 comes from converting the
linear change in $\reff$ to logarithmic and the factor of 1.2 derives from Equation~21 of
Sparks \& J\o rgensen (1993).

The slope of the FP in the $K$--band provides another equation of constraint, as its slope can
be affected by age and deviations from homology, but virtually not by metallicity.
From the fit to the Faber--Jackson (1976) relation in the $K$--band (\cite{pahre98kfp}), 
$\Delta\Ktot = 10.35 \Delta\sigma_0$, from Table~\ref{table-fpmodels-bc96-ubvrik}
$\Delta\Ktot ({\rm mag}) = +1.48 \Delta\log t - 0.16 \Delta$[Fe/H] (dex), so
the luminosity along the sequence varies as
\begin{equation}
\gamma = { +1.48 \Delta\log t - 0.16 \Delta {\rm~[Fe/H]~} ({\rm dex}) \over 10.35 \Delta\log\sigma_0 }
\label{fpmodels-eq-gamma}
\end{equation}
In this way the slope of the $K$--band FP (i.e., $a$ in $\reff \propto \sigma_0^{a}$; \cite{pahre98kfp}) 
provides the equation of constraint
\begin{equation}
1.528 \pm 0.083 = {1 \over 1+d} \left( { 2 \over 2 \gamma + 1 } \right)
\label{fpmodels-eq-kfp-constraint}
\end{equation}
where a new model parameter $d$ was introduced to represent the deviations of the family of 
ellipticals from a homologous family in dynamical structures.
In this notation, the mapping of velocity dispersions is 
$\log\sigma_0 = (1+d) \log\sigma_{\rm eff} + {\rm~const.}$ to provide a systematic variation of 
$\log\sigma_0 - \log\sigma_{\rm eff}$ along the elliptical galaxy sequence.
Introduction of the model parameter $d$ also allows for an additional equation of constraint from
the measurement of this mapping by Busarello \etal\ (1997), who found $d = 0.28 \pm 0.11$.

In summary, there are 22 equations of constraint represented by Equation~\ref{fpmodels-eq-color-change} 
from the comparisons of the optical and near--infrared FP between pairs of bandpasses 
(where $\Delta a_{j,i}$ are provided in Tables~\ref{table-fpmodels-kfp-optfp} and \ref{table-fpmodels-optfp-optfp}), and one 
equation of constraint each from the \mgtwosigma\ relation, the slope of the $K$--band FP relation,
and the dynamical non-homology measurement.
There are four free parameters:  (1) the variation in age $\Delta\log t$ from one end of the FP to 
the other; (2) the variation in metallicity $\Delta$[Fe/H] along the same sequence; 
(3) the size of the stellar populations gradient (equal for all elliptical galaxies), expressed for convenience
as a metallicity gradient, which produces a color gradient $\beta = \Delta ( \mu_i - \mu_j ) / \Delta \log r$; 
and (4) the size of the dynamical non-homology contribution $d$ to the mapping from $\sigma_0$ 
to $\sigma_{\rm eff}$.

\subsection{Solutions to the Physical Quantities in the Model for the Scaling Relations 
	\label{fpmodels-self-solutions} }

The variance was minimized orthogonal to the fit and the uncertainties for each measurement were included
in the construction of the Chi--squared statistic.
The uncertainties for the four measurements of the change in slope of the optical FP for the 
Prugniel \& Simien (1996) data set were intentionally doubled to account for the systematic effect that
these data do not explicitly account for the effects of color gradients.
The parameter $\Delta \log \sigma_0$ was set to $0.6$~dex to account for the range of velocity dispersion
occupied by nearly all the elliptical galaxies; this number merely scales up the model
parameters (except for $d$) without changing the significance of any parameter.
The least--squares solution was for the following values of the model parameters
\begin{equation}
\begin{array}{rcl}
\Delta \log t 					& = 	& +0.38 \pm 0.20 {\rm~dex} \\
\Delta {\rm~[Fe/H]} 				& = 	& +0.28 \pm 0.14 {\rm~dex} \\
{ \Delta {\rm~[Fe/H]} \over \Delta \log r }	& = 	& -0.26 \pm 0.28 {\rm~dex} \\
d 						& = 	& +0.17 \pm 0.10
\end{array}
\label{fpmodels-eq-model-bestfit1}
\end{equation}
The uncertainty estimates are taken from the covariance matrix.
The reduced Chi--square for this fit is close to unity at $1.33$, suggesting that 
the combination of the model and the uncertainty estimates in each of the observables
is a reasonable description of the properties of elliptical galaxies along their
sequence.

The results given in Equation~\ref{fpmodels-eq-model-bestfit1} for the first time describe the
underlying physical origin of the elliptical galaxy scaling relations using a self--consistent
model that accounts for population gradients, wavelength effects on the FP, systematic deviations
from homology, and a metal line--strength indicator.
The formal significance of the results in Equation~\ref{fpmodels-eq-model-bestfit1}, however, appear
to suffer from low significance for any given parameter:  $\Delta\log t$, $\Delta$[Fe/H], and
$d$ are all only significant at the 2~$\sigma$~CL, while the populations gradient is virtually
unconstrained and even consistent with zero.
There is significant correlation between the model parameters which is the underlying cause of
the reasonably large uncertainties on each parameter; the largest correlation coefficient is
$-0.7$ between $d$ and $\Delta\log t$, which is not surprising since either parameter (or a combination
of both) is essential for satisfying constraint from the slope of the near--infrared FP 
(Equation~\ref{fpmodels-eq-kfp-constraint}).

The inability to constrain the populations gradients should not be considered a problem, since
this model is actually only an indirect way of measuring populations gradients in ellipticals;
far better are direct measurements of \mgtwo\ line strength or color gradients.
For all of the following fits, an additional equation of constraint will be included to represent
the populations gradients:  $\Delta {\rm~[Fe/H]} / \Delta \log r = -0.22 \pm 0.01$~dex per 
decade of radius, as this is the mean of color and line--strength gradients from the 
literature in the analysis of Peletier (1993).
The least--squares solution then becomes:
\begin{equation}
\begin{array}{rcl}
\Delta \log t 					& = 	& +0.36 \pm 0.15 {\rm~dex} \\
\Delta {\rm~[Fe/H]} 				& = 	& +0.26 \pm 0.11 {\rm~dex} \\
{ \Delta {\rm~[Fe/H]} \over \Delta \log r }	&\equiv & -0.22 \pm 0.01 {\rm~dex} \\
d 						& = 	& +0.17 \pm 0.09
\end{array}
\label{fpmodels-eq-model-bestfit2}
\end{equation}
with $\chi^2/\nu = 1.27$.
When one or more parameters are set to zero, then the following series of 
solutions $(\Delta\log t,\Delta {\rm~[Fe/H]}, d, \chi^2/\nu)$ are obtained:  
\begin{equation}
\begin{array}{llll}
\Delta\log t = +0.73 	& \Delta {\rm~[Fe/H]} = 0 	& d = 0 	& \chi^2/\nu = 1.47 \\
\Delta\log t = 0 	& \Delta {\rm~[Fe/H]} = +0.50 	& d = 0 	& \chi^2/\nu = 3.20 \\
\Delta\log t = 0 	& \Delta {\rm~[Fe/H]} = 0 	& d = 0.30 	& \chi^2/\nu = 41 \\
\Delta\log t = +0.73 	& \Delta {\rm~[Fe/H]} = 0 	& d = 0.05 	& \chi^2/\nu = 1.50 \\
\Delta\log t = +0.58 	& \Delta {\rm~[Fe/H]} = +0.11 	& d = 0		& \chi^2/\nu = 1.44 \\
\Delta\log t = 0 	& \Delta {\rm~[Fe/H]} = +0.51	& d = 0.32 	& \chi^2/\nu = 1.52
\end{array}
\label{fpmodels-eq-settozero}
\end{equation}
In all cases, there is a significant or substantial increase in $\chi^2/\nu$ by factors between
1.13 to 40, suggesting that the full set of model parameters is required to 
provide an accurate representation of the observables.

\begin{figure}
	\epsscale{0.9}
	\plotone{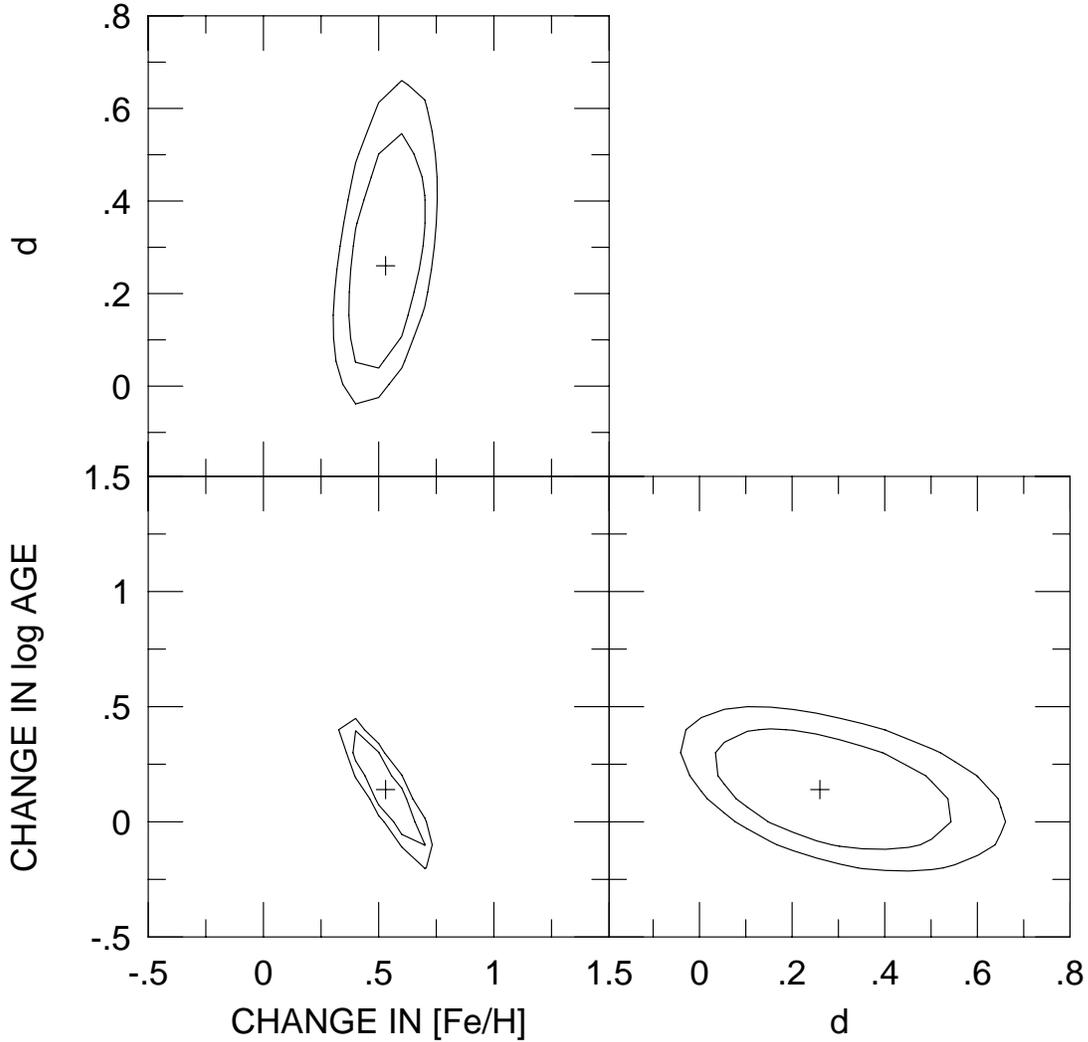}
	\caption[Contours of $\chi^2/\nu$ for the Self--Consistent Model for the FP]{
	Contour plots of $\chi^2/\nu$ for the self--consistent model described in the
	text using the stellar population synthesis models of Vazdekis \etal\ (1996) as
	an illustration of the joint confidence of pairs of parameters.
	In each figure, the $\chi^2$ minimum is identified, and two contours delimiting
	the 95\% and 99.7\% confidence regions are shown.
	While age and metallicity have a well--determined joint contribution specified by
	the ``3/2 rule'' (as is apparent by their $\chi^2/\nu$ valley with slope 3/2),
	the model parameter $d$, which describes the non-homology contribution,
	is poorly constrained by the observations.
	Since age and $d$ are jointly constrained by the slope of the $K$--band FP 
	(Equation~\ref{fpmodels-eq-model-bestfit3}), better independent constraints on 
	$d$ will further limit the allowed parameter
	space for age and thus further break the age--metallicity degeneracy.
	}
	\label{fig-fpmodels-model-chisq}
\end{figure}

Using the Vazdekis \etal\ (1996) models instead of the Bruzual \& Charlot (1996) models,
but still keeping $\Delta {\rm~[Fe/H]} / \Delta \log r = -0.22 \pm 0.01$~dex per 
decade of radius, produces the solution:
\begin{equation}
\begin{array}{rcl}
\Delta \log t 					& = 	& +0.14 \pm 0.07 {\rm~dex} \\
\Delta {\rm~[Fe/H]} 				& = 	& +0.53 \pm 0.05 {\rm~dex} \\
{ \Delta {\rm~[Fe/H]} \over \Delta \log r }	&\equiv & -0.22 \pm 0.01 {\rm~dex} \\
d 						& = 	& +0.26 \pm 0.07
\end{array}
\label{fpmodels-eq-model-bestfit3}
\end{equation}
with $\chi^2/\nu = 1.04$.
Since the model uncertainties have not been included in the $\chi^2$ statistic,
this reduction of 20\% in $\chi^2/\nu$ for the Vazdekis \etal\ (1996) models over the 
Bruzual \& Charlot (1996) models suggests that the former have a subtle improvement over
the latter in their treatment of the photometric properties of old stellar populations.
Contours of joint probability between pairs of the model parameters are plotted in
Figure~\ref{fig-fpmodels-model-chisq} for this solution.

\placefigure{fig-fpmodels-model-chisq}

If the differences in slope $\Delta a_{j,i}$ derived from the \Dnsigma\ relation are used 
instead of those from the quantity $\log\reff - 0.32 \meanmueff$, then the solution is
\begin{equation}
\begin{array}{rcl}
\Delta \log t 					& = 	& +0.16 \pm 0.09 {\rm~dex} \\
\Delta {\rm~[Fe/H]} 				& = 	& +0.50 \pm 0.05 {\rm~dex} \\
{ \Delta {\rm~[Fe/H]} \over \Delta \log r }	&\equiv & -0.22 \pm 0.01 {\rm~dex} \\
d 						& = 	& +0.25 \pm 0.11
\end{array}
\label{fpmodels-eq-model-bestfit4}
\end{equation}
with a much poorer $\chi^2/\nu = 3.04$.
The difference in $\chi^2/\nu$ between this solution (using the differences in $\Dn$) and the
previous solution (using the differences in $\log\reff - 0.32 \meanmueff$) can be directly
attributed to the significantly smaller uncertainties in the measurements of $\Delta a_{j,i}$
from the \Dnsigma\ relation in Tables~\ref{table-fpmodels-kfp-optfp} and \ref{table-fpmodels-optfp-optfp}.
The same effect in $\chi^2/\nu$ is found when the Bruzual \& Charlot models are used instead.
We suspect that these small formal uncertainties arise due to a poor sensitivity of the
$\Dn$ parameter to the subtle effects of color gradients, despite the apparent homogeneity and
repeatability in measuring this quantity.
On the other hand, the difference could point to overall limitations of the model if these small
uncertainties in $\Delta a_{j,i}$ are real.

\subsection{The Relative Roles of Various Constraints on the Model Solution 
	\label{fpmodels-self-relative-roles} }

Several remarks need to be made about the contribution of the various equations of constraint
towards the self--consistent solutions described above.
Broadband colors are notoriously poor at discriminating between age and metallicity effects,
which has been summarized elegantly by Worthey (1994) as the ``3/2 Rule'':  changes in
$\Delta \log t$ are virtually indistinguishable from changes in metallicity
$\Delta $[Fe/H]$ \approx {3 \over 2} \Delta \log t$.
Note that all solutions for this model have $\Delta \log t + {3 \over 2} \Delta $[Fe/H]$ \sim 0.75$~dex
(Bruzual \& Charlot models) or $0.95$~dex (Vazdekis \etal\ models).
The comparisons of the FP slopes in various optical and near--infrared bandpasses (represented 
by the $\Delta a_{j,i}$ terms) thus provide extremely good constraints on the joint contribution 
of age and metallicity to producing the slope of the FP at all bandpasses, but they do
not provide a unique discrimination between age and metallicity effects as the dominant cause
of the sequence.

A similar argument can be made as to the limitation of the \mgtwo\ index (in the \mgtwosigma\ relation)
in dealing with this age--metallicity degeneracy.
In our experience, however, this additional equation of constraint due to the \mgtwosigma\ relation
is essential to narrow the large parameter space that the populations gradients could occupy, 
since $\Delta$[Fe/H] and the color gradient $\beta$ enter into the $\Delta a_{j,i}$ equations in a 
fixed ratio {\sl for all colors} but in a different ratio for the \mgtwosigma\ relation.
Virtually all metal absorption line indices will have similar age--metallicity degeneracy
problems; the Balmer absorption lines of atomic hydrogen may not suffer the same problems, since 
these lines are quite sensitive to recent star formation activity.
Future modeling work along these lines could reduce this degeneracy by including Balmer line
measurements.

The introduction of the {\sl absolute} slope of the near--infrared FP as an additional equation of
constraint provides what is effectively a breaking of the age--metallicity degeneracy, since metallicity 
effects are unimportant at $K$ while age effects are significant.
The wrinkle caused by introducing the absolute slope of the near--infrared FP into the model is
that there can be an additional effect caused by deviations from dynamical homology which can, in part
or in whole, explain the deviation of the near--infrared FP from its virial expectation.
It was therefore necessary to introduce one more parameter to represent this dynamical non-homology,
and to include an additional equation of constraint governing it (as measured by \cite{busarello97}),
even though that constraint is not highly significant at the 2~$\sigma$~CL.
The large uncertainties in each model parameter in the simultaneous fit given by 
Equation~\ref{fpmodels-eq-model-bestfit2} can be directly traced back to the poor constraint provided
by the Busarello \etal\ (1997) measurement of $d$.
This is clearly the portion of the entire set of observables that needs substantial more work in the 
future in order to narrow the space occupied by all the model parameters.

\section{Discussion \label{fpmodels-discussion} }

\subsection{The Early--Type Galaxy Sequence   }

The global scaling relations provide a unique tool for investigating the underlying
physical properties which give rise to the sequence of elliptical galaxies.
While these correlations have significant and resolved intrinsic dispersion, they are
still quite thin and portray a remarkable homogeneity of galaxy properties from the $U$--band
to the $K$--band.

This homogeneity appears not to vary with the environment, implying a strong constraint
on possible variations in age between field and cluster early--type galaxies and contradicting
the predicted offset of 4~Gyr of semi--analytical, hierarchical galaxy formation models 
(\cite{kauffmann96}).
The current uncertainties in Galactic extinction correction dominate the random uncertainties 
of the observations at the level of $\sim 10$\% difference in age.
Other systematic effects, such as systematic variations in homology breaking or dark matter 
content between field and cluster early--type galaxies, are not constrained by the analysis
presented in this paper; future work to provide such constraints will necessarily rely upon
the larger uncertainties resulting from a distance--dependent parameterization of the 
observables.

The elliptical galaxy scaling relations in the near--infrared, with the exception of the 
$K$--band Faber--Jackson relation, do not follow the predictions of the virial theorem under
the assumptions of constant $M/L$ and homology.
This is an important clue as to the physical origins of these relations, which can immediately
exclude a number of simple models (\cite{pahre98kfp}) to explain the elliptical galaxy sequence.

The reduction of $\reff$ with increasing wavelength is an expected result of the presence of
stellar population gradients.
The fact that $\reff$ is a function of wavelength argues that any method of calculating intrinsic 
galaxy masses using the observables $\reff$ and $\sigma_0$ will be systematically flawed
(see the discussion in \cite{pahre98kfp}).
It is an open issue how best to match the observed effective radii, which are luminosity weighted
for a particular bandpass and hence affected by stellar populations gradients, to the half mass
radius of galaxies.
The latter quantity is the intrinsic property which is desired from the observations and readily
calculated in theoretical calculations, but its connection with any optical or 
near--infrared observations is still problematical.

The global properties of elliptical galaxies that are enumerated in \S\ref{fpmodels-general}
provide a large set of observables which should be accounted for by any viable model for the
physical properties which underlie and produce the elliptical galaxy sequence.
There are certainly more properties from X--ray, far--infrared, and radio wavelengths
which were not included in this list but ought to be in a more general discussion of the
fundamental nature of elliptical galaxies.

The parameter space occupied by variations in age and metallicity, the size of the mean populations 
gradients, and the deviations from a dynamically homologous family has been shown in 
\S\ref{fpmodels-selfconsistent} to be limited significantly by a large and homogeneous
sample of global optical and near--infrared photometric parameters and global spectroscopic 
parameters.
While the degeneracy of age and metallicity is difficult to overcome with such data,
observations in the metallicity insensitive $K$--band narrow the range of possible models
to age and/or dynamical non-homology causing the $K$--band FP slope, while still not excluding
metallicity as a contributor to the optical FP slope.
The explicit accounting for the effects of populations gradients on all relevant parameters, and
the inclusion of the slope of the \mgtwosigma\ relation, provide a further narrowing of the 
allowed parameter space within the model.

The modeling methodology that has been developed in \S\ref{fpmodels-selfconsistent} provides the
first self--consistent exploration of the underlying physical origins of the elliptical galaxy
scaling relations which can simultaneously account for the following observables:  
(1) the {\sl changes} of the slope of the FP among the $UBVRIK$ bandpasses; 
(2) the absolute value of the slope of the FP; 
(3) the effects of color gradients on the global properties of ellipticals; 
(4) the slope of the \mgtwosigma\ relation; 
and (5) the contribution of deviations from a dynamically homologous family to the slope of the FP.

The aperture color--magnitude relation\footnote{The distinction is made here between a \emph{global}
color--magnitude relation, for which the color is measured globally within the effective radius, and an
\emph{aperture} color--magnitude relation, for which the color is measured within a fixed, metric
aperture for all galaxies independent of their effective radii.  These two methods of measuring colors 
differ due to the presence of color gradients through Equation~\ref{fpmodels-eq-color-change}, such
that the aperture color--magnitude relation has a steeper slope.} has not been explicitly included
in the model developed in \S\ref{fpmodels-selfconsistent} since colors for the entire galaxy sample 
could not be measured in a self--consistent manner.
Nonetheless, a check on the model solutions in \S\ref{fpmodels-self-solutions} is to compare their predicted
slope for the aperture color--magnitude relation with the slope from observational data in the literature.
The solutions of Equations~\ref{fpmodels-eq-model-bestfit2} and \ref{fpmodels-eq-model-bestfit3},
where the former had the larger age spread and the latter had the larger metallicity spread, predict
slopes for the $(U-V)_0$ versus $\Vtot$ of $-0.11$ and $-0.09$, respectively, while the slope in the
Coma cluster is $-0.08 \pm 0.01$ (\cite{ble92b}), suggesting that the model with smaller a age spread is 
favored.
There is some uncertainty in this comparison, however, since $\partial (U-V)_0 / \partial \log t$
varies between the two models by 0.06~mag~dex$^{-1}$ and there might be small, additional systematic 
effects in matching the precise $U$--band filter used by Bower \etal\ (1992b).
Furthermore, taking the observed values of $\Delta a_{B,U}$ and $\Delta a_{V,B}$ from 
Table~\ref{table-fpmodels-optfp-optfp} and inserting their sum into Equation~\ref{fpmodels-eq-color-change}
produces an expected ratio of $(\Delta C_{U,V} - 1.875 \Delta\reffij ) / \Delta\log\sigma_0 = 0.56$,
which is similar to the value of $0.54$ in Bower \etal\ (1992b).

One difficult problem that can be posed both by the \mgtwo\ version of the near--infrared FP 
(\cite{pahre98kfp}) and the optical FP in J\o rgensen \etal\ (1996), is why this FP has much 
larger scatter than the standard $\sigma_0$ form of the FP.
While the \mgtwo\ index is an indicator of variations in metallicity (\cite{mould78}), 
it can be affected by ``filling''
due to a younger stellar component and it reflects the existence of stellar populations
gradients via \mgtwo\ line gradients (\cite{couture88}; \cite{gorgas90}; \cite{davies93}).
But \mgtwo\ does not reflect the intrinsic {\sl dynamical} effects which may vary along the elliptical
sequence---or even at any given point in the sequence.
Peletier (1993) argued that local velocity dispersion is not a universal predictor of \mgtwo; this
argument could be reversed to say that \mgtwo\ is not a universal predictor of velocity dispersion.
While the \mgtwo\ FP is not explicitly described by the model in \S\ref{fpmodels-selfconsistent},
its large scatter might reflect the real presence of a dynamical non-homology term $d > 0$ in the
FP that is not accounted for in the \mgtwo\ FP.

The model parameters derived in \S\ref{fpmodels-selfconsistent} imply that age and metallicity 
are varying along the early--type galaxy sequence in the sense that the most
luminous galaxies are the oldest and most metal rich.
If there exists a mass--age sequence among early--type galaxies, this might be inconsistent
with hierarchical models in which present day massive galaxies are built by successive
mergers of smaller, sub--galactic units (cf. \cite{kauffmann96}).
Furthermore, the sense of the metallicity variations is as expected if metallicity (and population
gradients in metallicity, see \S\ref{fpmodels-self-slope}) drives the color--magnitude 
relation (\cite{kodama97}) for elliptical galaxies.

The trend for more luminous galaxies to be more metal rich is in contradiction to the study of 
line indices of Trager (1997), who suggested that the most luminous galaxies are the oldest while
also being the most metal poor.
The correlation between age and metallicity in the Trager (1997) analysis, however, could be
caused at least in part by the correlated errors in the derived parameters.
Furthermore, there exists substantial scatter perpendicular to the correlation Trager proposes between
age and metallicity which cannot be explained by correlated errors---this perpendicular scatter is
exactly in the sense that age and metallicity are proposed to correlate in the present paper.
Finally, the comparison of $K$--band surface brightness fluctuations measurements with $(V-I)$
color for 11 galaxies (\cite{jensen97}) also suggests that age variations of a factor of two to
three are occurring along the elliptical galaxy sequence, further contradicting the large age spreads
of Trager (1997).

\subsection{Homology Breaking and Variations in Dark Matter Content   }

The analysis in this paper shows that a wavelength--independent effect, such as systematic
homology breaking or variations in dark matter content, contributes to the slope of the FP
at all wavelengths.
While hints of this effect were found in several previous analyses (\cite{prugniel96}; 
\cite{pahre97stromlo}), the  first study included unusual assumptions about the form of the
FP (see \S\ref{fpmodels-opt-ir-new} above) and no change in $\reff$ with wavelength, while
both suffered from potential distant--dependent biases.
Furthermore, the present paper re--analyzed data from many different literature sources at
many wavelengths to demonstrate that the effect is present in all such data.
It is an important result that the assumption of homology among the family of early--type 
galaxies may not entirely correct.
It is also possible that the dark matter content among this family of galaxies varies 
systematically with mass; only deep, spectral observations and detailed analysis of a 
large sample of early--type galaxies can determine conclusively if this effect might be
due to variations in dark matter content.
The observed measurement of $d = 0.28 \pm 0.11$ by Busarello \etal\ (1997), and the
identification of a similar property in the numerical simulations of dissipationless merging
by Capelato \etal\ (1995), suggests that dynamical homology breaking is probably more
likely than dark matter variations as the cause of this wavelength independent effect.

Deviations of the model parameter $d$ from zero were constructed to portray the effects of dynamical
non-homology on the slope of the FP via the mapping from $\sigma_0$ to $\sigma_{\rm eff}$, but it is
important to consider if the non-homology represented by $d > 0$ could be a result of structural
non-homology.
Graham \& Colless (1997) showed that the effects on the FP are minimal for the breaking of structural
homology, but this result may not be conclusive since distance (such as the resolved depth of
the cluster) could systematically affect their Virgo cluster data, thereby hiding the structural homology
breaking.
The fundamental problem with invoking structural non-homology, as they pointed out, is that any increase 
in $\reff$ is compensated by a decrease in $\meanmueff$ (they actually found a slight over-compensation),
which effectively nulls the result.
This is basically a different way of thinking about the fact that only small uncertainties enter the 
FP through the quantity $\log\reff - 0.32 \meanmueff$.
In summary, since changes in $\meanmueff$ virtually compensate for changes in $\reff$ in the FP,
there is no significant way that values of the non-homology parameter $d > 0$ can be
traced back to systematic mismeasures of $\reff$.

\subsection{Evolution of the Form of the Early--Type Galaxy Correlations  }

Since the derived model solutions in \S\ref{fpmodels-self-solutions} have small variations in age
along the early--type galaxy sequence, the slope of the color--magnitude relation is expected to evolve
slowly with redshift.
The model with a total age variation of $0.14$~dex (Equation~\ref{fpmodels-eq-model-bestfit3}) 
predicts that the slope of the color--magnitude relation
should increase by $0.01$ by $z = 0.5$, which does not contradict the comparison of the observations of 
Bower \etal\ (1992b) at $z=0$ and Ellis \etal\ (1997) at $z=0.5$, especially considering the potential
systematic errors associated with rejecting the lower luminosity outliers in the higher redshift data.
The slope of the color--magnitude relation is measured to an accuracy of only 0.02--0.04 by
Stanford, Eisenhardt, \& Dickinson (1998) for 17 clusters at $0.3 < z < 0.9$, so a change of 0.01 in the
slope to $z=0.5$ could certainly exist, especially considering the small systematic uncertainties which could
result from the variations in the rest--frame wavelengths sampled for each cluster.
The model with a larger age variation of $0.36$~dex (Equation~\ref{fpmodels-eq-model-bestfit2}), 
however, has a larger predicted change in the slope of the color--magnitude relation and may be 
marginally in conflict with those data.
Direct visual inspection of the color--magnitude relations (``blue--K'') in Stanford \etal\ for all
clusters at $z > 0.55$, however, leaves the distinct impression that larger variations in the slope
of the color--magnitude relation are allowed by the data (with the one exception being GHO~1603+4313).
Furthermore, the large aperture sizes used by Stanford \etal\ for measuring colors reduces the predicted
evolution of the slope of the color--magnitude relation due to the effects of color gradients in the larger 
galaxies.
As a result, both the model solutions derived in \S\ref{fpmodels-self-solutions} are probably consistent
with the slope of the color--magnitude relation at intermediate redshifts.

\begin{figure}
	\epsscale{1.0}
	\plotone{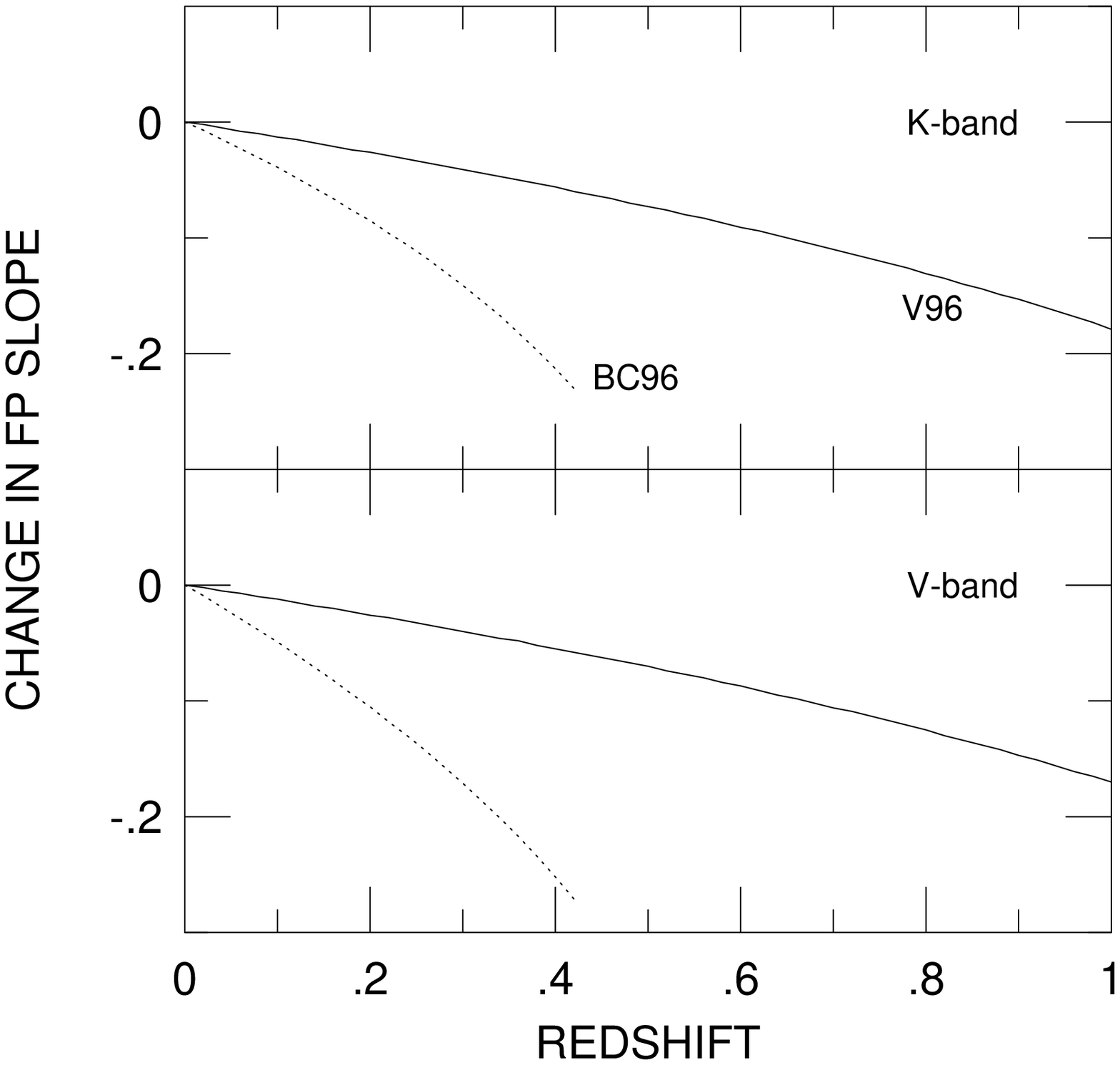}
	\caption[The Evolution of the FP With Redshift for the Self--Consistent Model Describing the FP]{
	The evolution of the FP with redshift for the two self--consistent model solutions 
	(\cite{vazdekis96}, V96, Equation~\ref{fpmodels-eq-model-bestfit3}; 
	\cite{bc96}, BC96, Equation~\ref{fpmodels-eq-model-bestfit2})
	which describe the intrinsic physical properties of the early--type galaxy sequence.
	The cosmology assumed is $(H_0,\Omega_0,\Lambda_0)=(75,0.2,0)$, and the oldest galaxies are
	taken to be the age of the universe in the present day.
	The models are arbitrarily cutoff when the youngest galaxies in the early--type galaxy sequence
	reach an age $< 1$~Gyr.
	The V96 model solution predicts a present--day age spread of 35\% along the sequence, while the
	BC96 model solution predicts a present--day age spread of a factor of two.
	}
	\label{fig-fpmodels-fpevol}
\end{figure}

The model constructed in \S\ref{fpmodels-selfconsistent} also makes specific predictions about the
behavior of the FP relations with look-back time or redshift.
Age appears to be a significant contributor to the slope of the FP, in the sense that the most
luminous elliptical galaxies might be as much as twice as old as the least luminous galaxies.
In this model, the slope of the FP should evolve with redshift in the sense that the slope $a$
in $\reff \propto \sigma_0^{a}$ will {\sl decrease} with redshift, since the youngest galaxies at one
end of the FP will evolve more quickly than the oldest galaxies at the other end.
The specific predictions for the solutions from the 
Bruzual \& Charlot models (Equation~\ref{fpmodels-eq-model-bestfit2}) and the
Vazdekis \etal\ models (Equation~\ref{fpmodels-eq-model-bestfit3}) are shown in Figure~\ref{fig-fpmodels-fpevol}.
The evolution of the slope of the FP due to the presence of a dynamical non-homology effect is more
complicated.
Numerical simulations of dissipationless merging (\cite{capelato97}) seem to suggest that second and later
generation mergers produce a slightly steeper FP than the first generation mergers.
Since viewing the FP slope at larger redshifts would then be looking at earlier generations of mergers,
their interpretation leads to the prediction that the role of dynamical non-homology should increase 
with redshift, thereby increasing $d$ and decreasing the slope $a$ with redshift at all wavelengths.
This prediction should be treated with caution, however, since it is not clear that the origin of dynamical
non-homology effects are in the merging process studied in the numerical simulations.

\placefigure{fig-fpmodels-fpevol}

\subsection{Limitations of This Approach   }

The obvious limitation of this approach is that it still is an empirical description of the
observations, not a theoretical construct based on first principles and galaxy formation
theory.
Nonetheless, it is a first step towards providing detailed, quantitative constraints on the
properties that any viable theoretical model for galaxy formation and evolution needs to reproduce.

The only observed property of elliptical galaxies that is not explicitly described by this
model is the super--solar enrichment of Mg relative to Fe, although it could certainly be accommodated
by the inclusion of recent studies of $\langle$Fe$\rangle$ (such as by \cite{jorgensen98} or 
\cite{trager98}) and attempts to model this enrichment (\cite{weiss95}).
It may be important to reanalyze the galactic wind models (\cite{arimoto87}) with these recent
super metal--rich, $\alpha$--element enhanced, stellar populations models in a way which embodies
the ongoing research into the relative contributions of Type~Ia and II supernovae to the chemical
enrichment of the host galaxies.
This problem is not just a challenge for stellar populations synthesis models, but also
for supernova nucleosynthesis, galaxy formation models, and galactic wind models.

A more subtle limitation of this approach is its indirect inclusion of stellar populations gradients,
such that the size of these gradients is virtually unconstrained by the model.
Clearly, the optimal method of constraining stellar populations gradients is by directly observing
them in various colors and line strengths.
It is truly surprising that very few new observations have been reported since the review
of Peletier (1993), despite the advent of large--format CCD and IR arrays, a sky--subtraction
independent parameterization (\cite{sparks93}), and the wealth of photometry that has been
obtained from the $U$ to the $K$ bandpasses (\cite{ble92a}; \cite{jfk95a}; \cite{smith97}; 
\cite{lucey97}; \cite{pahre98kfpdata}).
Furthermore, comparing color gradient ratios (between different colors) will provide a strong
tool to discriminate between stellar populations gradients and a diffuse component of dust
(\cite{wise96}).

\section{Summary  \label{fpmodels-summary} }

The Fundamental Plane slope has been shown to steepen in a systematic way from shorter to longer 
wavelengths.
The methodology presented here shows that changes of the FP slope between bandpasses can 
be measured accurately by a distance independent construction of the observables.
This method is robust and typically reduces the uncertainty of the comparison by a factor of two, 
thereby allowing for more detailed model comparisons.

This paper presents for the first time a comprehensive model of the changes in global properties
of elliptical galaxies that simultaneously accounts for a wide range of observables, namely:
(1) the changes in slope of the FP between bandpasses; (2) the slope of the near--infrared FP;
(3) the slope of the \mgtwosigma\ relation; (4) the presence and effects of stellar 
populations gradients; and (5) the presence of systematic deviations of the internal dynamical
structures of elliptical galaxies from a homologous family.
The observational constraints imposed by the last element of this model is clearly the weakest
point and should be substantially improved upon in the future by obtaining velocity dispersion 
profiles for large samples of galaxies, such as in a rich cluster.
Due to this observational shortcoming, this model does not yet provide highly significant
measurements of the individual model parameters defining the variations in age and metallicity
from one end of the FP to the other.
The model, however, does provide a framework to re-evaluate these parameters as soon as newer 
and higher quality data become available.

\acknowledgments

This research has made use of the NASA/IPAC Extragalactic Database (NED)
which is operated by the Jet Propulsion Laboratory, California Institute
of Technology, under contract with the National Aeronautics and Space
Administration.
During the course of this project, M.~A.~P. received financial support from Jesse
Greenstein and Kingsley Fellowships, and Hubble Fellowship grant 
HF-01099.01-97A from STScI (which is operated by AURA under NASA 
contract NAS5-26555).
S.~G.~D. was supported in part by grants from the NSF (AST--9157412) and the
Bressler Foundation.


\clearpage


\makeatletter
\def\jnl@aj{AJ}
\ifx\revtex@jnl\jnl@aj\let\tablebreak=\nl\fi
\makeatother
\scriptsize
\begin{deluxetable}{lcccccccccccccc}
\tablewidth{0pc}
\tablecaption{Comparison of the Slope of the FP in the Optical and Near--Infrared
	\label{table-fpmodels-kfp-optfp} }
\tablehead{
Literature		&$\lambda$ & \multicolumn{6}{c}{Fundamental Plane}	&& \multicolumn{6}{c}{ \Dnsigma\ } \\
\cline{3-8} \cline{10-15}
Source			&	   & $\Delta a$ &$\pm$& $c_{K,opt}$ &$\pm$& rms &N	 && $\Delta a$ & $\pm$ & $c_{K,opt}$ & $\pm$ & rms  & N \\
			&	   &	     &     &     &     & (dex)& &        &&            &       &     &       & (dex)
}

\startdata

All Data		&$V$\tablenotemark{a} & 0.18 &  0.03 &  0.60 &  0.07 &  0.05 & 239 && 0.18 &  0.02 & -0.41 &  0.06 &  0.04 & 249 \nl
\nl
J\o rgensen \etal\ (1996) & $U$	   & 0.51 &  0.13 &  0.28 &  0.30 &  0.03 &  20 && 0.38 &  0.05 & -0.88 &  0.10 &  0.03 &  20 \nl
J\o rgensen \etal\ (1996) & $B$	   & 0.32 &  0.08 &  0.57 &  0.17 &  0.03 &  25 && 0.30 &  0.08 & -0.70 &  0.18 &  0.03 &  26 \nl
Faber \etal\ (1989)	& $B$	   & 0.19 &  0.04 &  0.91 &  0.10 &  0.06 & 145 && 0.17 &  0.05 & -0.37 &  0.12 &  0.05 & 149 \nl
J\o rgensen \etal\ (1996) & $g$	   & 0.21 &  0.12 &  0.59 &  0.28 &  0.04 &  26 && 0.23 &  0.06 & -0.52 &  0.15 &  0.03 &  27 \nl
Lucey \etal\tablenotemark{b}	& $V$	   & 0.23 &  0.04 &  0.50 &  0.10 &  0.05 &  83 && 0.21 &  0.02 & -0.48 &  0.05 &  0.03 & 135 \nl
J\o rgensen \etal\ (1996) & $r$	   & 0.17 &  0.06 &  0.54 &  0.14 &  0.04 &  55 && 0.12 &  0.02 & -0.28 &  0.06 &  0.03 &  56 \nl
Smith \etal\ (1997)	& \rkc	   & 0.17 &  0.07 &  0.45 &  0.16 &  0.05 &  44 && 0.14 &  0.02 & -0.32 &  0.04 &  0.02 &  44 \nl
Scodeggio \etal\ (1997)	& \ikc	   & 0.22 &  0.10 &  0.04 &  0.24 &  0.04 &  43 && \nodata  &\nodata  &\nodata  &\nodata  &\nodata  &\nodata  \nl

\tablenotetext{a}{$^\dag$The complete sample uses data from $B$ to $I_{\rm C}$ bandpasses that have
 	been converted to $V$ assuming mean colors, except for the data of Faber \etal\ (1989)
 	which have been converted from $B$ to $V$ using their measurements of $(B-V)$.}
\tablenotetext{b}{$^\ddag$Lucey \etal\ refers to the combined sample of Lucey \& Carter (1988),
 	Lucey \etal\ (1991a,b), and Lucey \etal\ (1997).}

\enddata
\end{deluxetable}
\normalsize



\makeatletter
\def\jnl@aj{AJ}
\ifx\revtex@jnl\jnl@aj\let\tablebreak=\nl\fi
\makeatother
\scriptsize
\begin{deluxetable}{lcccccccccccccccc}
\tablewidth{0pc}
\tablecaption{Comparison of the Slope of the FP Among Various Optical Bandpasses
	\label{table-fpmodels-optfp-optfp} }
\tablehead{
Literature &$\lambda_1$	& Literature &$\lambda_2$ & \multicolumn{6}{c}{Fundamental Plane}	&& \multicolumn{6}{c}{ \Dnsigma\ } \\
\cline{5-10} \cline{12-17}
Source 1   &            & Source 2   &            & $\Delta a$ &$\pm$& $c_{\lambda_1,\lambda_2}$ &$\pm$& rms &N	
	&& $\Delta a$ & $\pm$ & $c_{\lambda_1,\lambda_2}$ & $\pm$ & rms  & N \\
& &		&	   &	     &     &     &     & (dex)& &        &&            &       &     &       & (dex)
}

\startdata

JFK96      & $r$        & JFK96      & $U$ & 0.22 &  0.03 &  0.01 &  0.07 &  0.03 &  45 && 0.17 &  0.03 & -0.42 &  0.08 &  0.02 &  45 \nl
JFK96      & $B$        & JFK96      & $U$ & 0.13 &  0.04 & -0.15 &  0.09 &  0.02 &  46 && 0.09 &  0.03 & -0.22 &  0.07 &  0.02 &  46 \nl
P96	   & $B$	& P96	     & $U$ & 0.13 &  0.01 & -0.13 &  0.02 &  0.03 & 353 && \nodata &\nodata &\nodata &\nodata &\nodata &\nodata \nl
P96	   & $V$	& P96	     & $B$ & 0.05 &  0.01 &  0.19 &  0.01 &  0.02 & 406 && \nodata &\nodata &\nodata &\nodata &\nodata &\nodata \nl
JFK96      & $r$        & JFK96      & $B$ & 0.18 &  0.04 & -0.06 &  0.10 &  0.03 &  36 && 0.10 &  0.04 & -0.23 &  0.08 &  0.02 &  37 \nl
JFK96      & $r$        & F89        & $B$ & 0.18 &  0.08 &  0.00 &  0.18 &  0.09 &  50 && 0.06 &  0.04 & -0.13 &  0.09 &  0.08 &  52 \nl
JFK96      & $r$        & JFK96      & $g$ & 0.08 &  0.02 & -0.04 &  0.05 &  0.02 &  79 && 0.09 &  0.01 & -0.20 &  0.03 &  0.02 &  80 \nl
JFK96      & $r$        & L91/97     & $V$ & 0.02 &  0.03 &  0.05 &  0.07 &  0.02 &  54 && 0.03 &  0.02 & -0.04 &  0.05 &  0.02 &  73 \nl
P96	   & \rkc	& P96	     & $V$ & 0.03 &  0.01 &  0.10 &  0.02 &  0.01 & 256 && \nodata &\nodata &\nodata &\nodata &\nodata &\nodata \nl
Smi97      & \rkc       & L91/97     & $V$ & 0.00 &  0.04 &  0.18 &  0.08 &  0.02 &  23 && 0.01 &  0.02 & -0.02 &  0.04 &  0.01 &  24 \nl
Sco97      & \ikc       & L91/97     & $V$ & 0.10 &  0.06 &  0.25 &  0.14 &  0.04 &  61 && \nodata &\nodata &\nodata &\nodata &\nodata &\nodata \nl
P96	   & \ikc	& P96	     &\rkc & 0.06 &  0.01 &  0.24 &  0.03 &  0.02 & 256 && \nodata &\nodata &\nodata &\nodata &\nodata &\nodata \nl
Sco97      & \ikc       & JFK96      & $r$ & 0.17 &  0.07 & -0.00 &  0.15 &  0.04 &  45 && \nodata &\nodata &\nodata &\nodata &\nodata &\nodata \nl

\tablerefs{
The literature sources referenced in this table are as follows.
F89:  Faber \etal\ (1989).
L91/97:  Lucey \& Carter (1988), Lucey \etal\ (1991a,b), and Lucey \etal\ (1997)
J96:  J\o rgensen \etal\ (1996).
P96:  Prugniel \& Simien (1996).
Smi97:  Smith \etal\ (1997).
Sco97:  Scodeggio \etal\ (1997).
}

\enddata
\end{deluxetable}
\normalsize



\makeatletter
\def\jnl@aj{AJ}
\ifx\revtex@jnl\jnl@aj\let\tablebreak=\nl\fi
\makeatother
\small
\begin{deluxetable}{lccccc}
\tablewidth{0pc}
\tablecaption{Variations in Magnitude for Various Bandpasses for the Bruzual \& Charlot (1996) and
		Vazdekis \etal\ (1996) Models
	\label{table-fpmodels-bc96-ubvrik} }
\tablehead{
& \multicolumn{2}{c}{Bruzual \& Charlot (1996)} && \multicolumn{2}{c}{Vazdekis \etal\ (1996)} \\
\cline{2-3} \cline{5-6}
Bandpass	& $\partial m / \partial \log t$ & $\partial m / \partial $[Fe/H] 
	& & $\partial m / \partial \log t$ & $\partial m / \partial $[Fe/H]	\\
		& (mag~dex$^{-1}$)		 & (mag~dex$^{-1}$) 
		&& (mag~dex$^{-1}$)		 & (mag~dex$^{-1}$)
}

\startdata

$U$		& $+2.934$	& $+1.79$ 	 && $+2.480$  & $+1.50$  \nl
$B$		& $+2.428$	& $+1.12$ 	 && $+2.010$  & $+1.04$  \nl
$V$		& $+2.165$	& $+0.86$ 	 && $+1.755$  & $+0.78$  \nl
\rkc\		& $+2.033$	& $+0.73$ 	 && $+1.629$  & $+0.63$  \nl
\ikc\		& $+1.925$	& $+0.65$ 	 && $+1.540$  & $+0.45$  \nl
$K$		& $+1.480$	& $-0.16$ 	 && $+1.566$  & $-0.13$  \nl
Mg$_2$		& $+0.174$	& $+0.278$ 	 && $+0.119$  & $+0.199$  \nl

\enddata
\end{deluxetable}
\normalsize

\clearpage



\clearpage

\end{document}